\documentclass[11pt]{article}
\usepackage{amsmath, latexsym,ifthen, amsfonts,amsbsy}
\usepackage{setspace,subfig}
\usepackage{natbib}
\usepackage{relsize}
\renewcommand{\harvardurl}[1]{\textbf{URL:} \url{#1}}
\usepackage{booktabs}
\usepackage{tkz-graph}
\usepackage{pgf}
\usepackage{setspace}
\usepackage{times}
\usepackage{lipsum}
\usepackage{subfig}
\usepackage{tabularx}
   
\usepackage{amsmath,amsthm,amssymb,ifthen}
\usepackage{graphicx}
\usepackage[mathscr]{eucal}

\usepackage{tikz}
\usetikzlibrary{arrows,shapes.arrows,shapes.geometric,
	shapes.multipartb,backgrounds,decorations.pathmorphing,positioning,fit,automata}
\tikzset{
	>=stealth',
	true/.style={
		rectangle,
		draw=black, very thick,
		text width=6.5em,
		minimum height=2em,
		text centered,
		fill=gray, opacity = 0.5},
	punkt/.style={
		rectangle,
		rounded corners,
		draw=black, very thick,
		text width=6.5em,
		minimum height=2em,
		text centered},
	est/.style={
		circle,
		draw=black, very thick,
		text centered},
	shade/.style={
		circle,
		draw=black, very thick, fill=gray!50,
		text centered},
	weight/.style={
		circle,
		draw=black, very thick,
		text width=6.5em,
		minimum height=2em,
		text centered},
	pil/.style={
		->,
		thick,
		shorten <=2pt,
		shorten >=2pt,},
	double/.style={
		<->,
		thick,
		shorten <=2pt,
		shorten >=2pt,},
	dash/.style={
		dashed,
		thick,
		shorten <=2pt,
		shorten >=2pt,},
	dashdouble/.style={
		<->,
		dashed,
		thick,
		shorten <=2pt,
		shorten >=2pt,}
}
\usepackage{enumerate}

\usepackage{etex}
\usepackage{booktabs}
\usepackage{amsfonts}
\usepackage[latin1]{inputenc}
\usepackage{soul,color}
\usepackage{chngcntr}
\usepackage{amsthm}
\usepackage{verbatim}
\usepackage[hidelinks]{hyperref}
\usepackage{changepage}
\usepackage{amssymb}
\usepackage{adjustbox}

\usepackage{caption} 
\usepackage{xfrac}
\usepackage{graphicx}
\newcommand{\ind}{\rotatebox[origin=c]{90}{$\models$}}

\newtheoremstyle{note}% <name>
{8pt}% <Space above>
{8pt}% <Space below>
{}% <Body font>
{}% <Indent amount>
{\bfseries}% <Theorem head font>
{:}% <Punctuation after theorem head>
{.5em}% <Space after theorem headi>
{}% <Theorem head spec (can be left empty, meaning `normal')>

\theoremstyle{note}
\newtheorem{theorem}{Theorem}

\newtheorem{remark}{Remark}
\usepackage[T1]{fontenc}

\newtheorem{proposition}{Proposition}
\allowdisplaybreaks
\newcommand\numberthis{\addtocounter{equation}{1}\tag{\theequation}}
\usepackage{multirow}
\usepackage[left=1in,right=1in,top=1in,bottom=1in]{geometry}
\usepackage{algorithm}
\date{}
\usepackage{tikz}

\usepackage{soulpos}
\soulregister\cite7
\soulregister\citep7
\soulregister\ref7
\soulregister\emph7
\soulregister\eqref7
\soulregister\pageref7
\definecolor{mygreen}{RGB}{144,241,47}

\newcommand{\tred}{\textcolor{red}}

\newcommand{\nind}{\not\!\perp\!\!\!\perp}
\newcommand{\m}{\mathcal{M}}

\newcommand{\Mu}{\mathcal{M}_{\text{\it union}}}

\def\bSig\mathbf{\Sigma}

\usepackage[figuresright]{rotating}

\usepackage{authblk}
\author[]{Linbo Wang\thanks{Address for correspondence: Linbo Wang, Department of Biostatistics, Harvard School of Public Health, 677 Huntington Avenue, Boston, Massachusetts 02115 \\
		Email: linbowang@g.harvard.edu} }
\author[]{Eric Tchetgen Tchetgen}
\affil[]{Harvard T.H. Chan School of Public Health, Boston, Massachusetts, U.S.A.}
	{\raise0.5ex\hbox{$#1$}\! \left/ \! \lower0.5ex\hbox{$#2$}\right.}
	
\begin{document}

%	\title{Identification and estimation of average causal effects using instrumental variables}
	
\title{Bounded, efficient and multiply robust  estimation of average treatment effects using instrumental variables}

	\clearpage \maketitle
	%  This label and the label ``lastpage'' are used by the \pagerange
	%  command above to give the page range for the article
	
%	\tableofcontents

\begin{abstract}	
	{	
		Instrumental variables (IVs) are widely used for estimating causal effects in the presence of unmeasured confounding. Under the standard IV model, however, the average treatment effect (ATE) is only partially identifiable. To address this, we propose novel assumptions that allow for identification of the ATE. Our identification assumptions are clearly separated from model assumptions needed for estimation, so that researchers are not required to commit to a specific observed data model in establishing identification. We then construct multiple estimators that are consistent under three different observed data models, and multiply robust estimators that are consistent in the union of these observed data models. We pay special attention to the case of binary outcomes, for which we obtain bounded estimators of the ATE that are guaranteed to lie between -1 and 1. Our approaches are illustrated with simulations and a data analysis evaluating the causal effect of education on earnings.
	}
\end{abstract}	

\section{Introduction}

Observational studies are often used to infer  treatment effects in social and biomedical sciences. In these studies, the treatment assignment may be associated with various background variables that are associated with the outcome, causing the unadjusted  treatment effect estimate to be biased. These background variables are often called confounders. A major challenge of causal inference in observational studies is that in practice, these confounding variables are often not fully observed, making it impossible to identify the treatment effect in view. In such settings, instrumental variable (IV) methods are useful in dealing with unmeasured confounding and have gained popularity among econometricians, statisticians and epidemiologists. Intuitively, conditional on baseline covariates, a valid IV affects the outcome through its effect on the treatment but is otherwise unrelated to the outcome. 

Traditionally, the IV methods have often been used to identify and estimate the parameters indexing a system of linear  structural equation models (SEMs)  \citep{wright1928tariff,goldberger1972structural}; see \citet[][\S 18.4.1]{wooldridge2010econometric} and \cite{clarke2012instrumental} for recent reviews.  Under correct specification of the linear SEMs,  the population average treatment effect (ATE) is equal to a certain parameter in the SEMs and can thus be consistently estimated. 
One such SEM can be inferred from the following system of linear regression models:
 \begin{subequations}
 	\begin{flalign*}
 	D &= \alpha_0 + \alpha_1 Z + \alpha_2 X + \alpha_3 U + \epsilon_D, \numberthis \label{eqn:lsem1} \\
 	Y &= \beta_0 + \beta_1 D + \beta_2 X + \beta_3 U + \epsilon_Y, \numberthis \label{eqn:lsem2}
 	\end{flalign*}
 \end{subequations}
 where $Z$ is an instrumental variable, $D$ is a continuous treatment, $Y$ is a continuous outcome, $X$ and $U$ denote observed and unobserved baseline covariates, respectively, $Z\ind U \mid X$ and the error terms are independent: $\epsilon_D \ind \epsilon_Y$. 
In models \eqref{eqn:lsem1} and \eqref{eqn:lsem2}, $\alpha_1\neq 0$ such that $Z$ is associated with the treatment, and  the fact that conditional on $D,X$ and $U$, $Z$ is excluded from the outcome model encodes the assumption that $Z$ has no direct effect on $Y$.
One can then show the parameter $\beta_1$ is identifiable and equals the ATE. Furthermore, it can be consistently estimated via
%  the Wald estimator \citep{wald1940fitting} in the case of binary instruments or 
two-stage least squares \citep{theil1953repeated}, in which one first obtains  estimates of $E[D\mid Z,X]$ through a linear regression, and then  regresses $Y$ on $\hat{E}[D\mid Z,X]$ and $X$ to get the treatment effect estimate $\hat{\beta}_1$.
However,  \eqref{eqn:lsem1} and \eqref{eqn:lsem2} impose strong parametric assumptions on the underlying data generating process. 
% in particular, they assume that the treatment effect is constant among all individuals in the population, which may be biologically impossible for applications in the health sciences,  especially in the case of binary outcomes \citep[e.g.][]{angrist1996identification}. 
Moreover, a fundamental limitation with relying on models like \eqref{eqn:lsem1} and \eqref{eqn:lsem2} is that they impose one set of assumptions, which conflates the definition, identification and estimation of the treatment effect.  For example, the target parameter $\beta_1$ may not even be well-defined if model \eqref{eqn:lsem2} is misspecified.

More recently, starting with the seminal work of \cite{imbens1994identification}, \cite{baker1994paired} and \cite{angrist1996identification}, more attention has been drawn to using the IV model to estimate the local average treatment effect (LATE) \citep[e.g.][]{abadie2002instrumental,abadie2003semiparametric,tan2006regression,cheng2009efficient,ogburn2015doubly},  defined as the average treatment effect for the so-called complier subgroup,
partly because it is nonparametrically defined and can be identified under an assumption that the effect of the IV on the treatment is not confounded and  a certain monotonicity assumption described in Section \ref{sec:framework}.
%Under the monotonicity assumption that the instrument is always good for the treatment, it is shown that the LATE equals the Wald estimand  \citep{imbens1994identification,tan2006regression}
%$$ 
%	\dfrac{E_XE[Y\mid Z=1,X] - E_XE[Y\mid Z=0,X]}{E_XE[D\mid Z=1,X] - E_XE[D\mid Z=0,X]}.
%$$
In this approach, assumptions needed for identification are usually clearly separated from assumptions needed for estimation, so that a researcher is not required to commit to  a specific observed data model in establishing identification. One may then  construct multiple estimators that are consistent under different
observed data models, or even estimators that are \emph{doubly robust} in the sense that they are consistent in
the union of two observed data models. 
However, the LATE concerns an unknown subset of the population which may be highly selective: the compliers may be more likely to
believe that they would benefit from the treatment \citep[e.g.][]{robins1996identification}. Furthermore, the definition of LATE depends on the particular IV that is available \citep[][p. 605]{wooldridge2010econometric}.
As a result, the LATE would in general differ from the ATE, which is arguably the causal parameter of interest in most observational studies \citep{imbens2010better}. We refer interested readers to  \cite{deaton2009instruments}, \cite{heckman2010comparing}, \cite{pearl2011principal} and \cite{aronow2013beyond} for additional discussions on whether the LATE is  of genuine scientific interest.

To the best of our knowledge, to date there has only been limited work focusing on ATE in the context of an IV, while successfully maintaining  identification and estimation assumptions clearly separated. 
Prior to our work,  \cite{angrist2013extrapolate} and \cite{aronow2013beyond} assume that conditional on a covariate profile, the ATE is the same as the LATE, so that  it can be identified under a monotonicity assumption, provided that the causal effect of the IV on the treatment is not confounded. 
\cite{hernan2006instruments} and \cite{vansteelandt2015robustness} instead assume  absence of treatment effect modification by the instrument among the treated
and untreated population, respectively. 
%show that the ATE is identifiable under the assumption that for all subjects, treatment and unmeasured confounders do no interact on the additive scale to cause the outcome. 
% treatment  homogeneity assumption of no treatment effect modification by the instrument among the treated and untreated population, respectively. 
%Their assumption is guaranteed to hold under the null of no treatment effect.
%, but can be too strong otherwise as it is an individual-level assumption.
% but can be 
%questionable otherwise as the treatment effects are likely to differ between the subpopulations who take treatment (or do not take treatment) under different values of the instruments  \citep{tan2010marginal}. 
In this paper, we propose two alternative no-interaction assumptions involving the unobserved confounders that allow for identification of the ATE. 
 Our first assumption is a  generalization of  linear model \eqref{eqn:lsem1};  our second assumption 
% is a generalization of \cite{hernan2006instruments}'s.
 is similar to \cite{hernan2006instruments}'s treatment homogeneity assumption in that it is guaranteed to hold under the null of no treatment effect, but it applies to the whole population rather than the treated (or untreated) population. 
We also do not rely on the monotonicity assumption for identification, and allow for  instruments that are confounded with the treatment.  One interesting observation is that under both of our identification assumptions,  the ATE can be represented by the same observed data functional so that in the estimation stage, we can target  a single statistical parameter. This parameter is called the average Wald estimand, a generalization of the Wald estimand \citep{wald1940fitting} to accommodate baseline covariates $X$.
 By carefully parameterizing the efficient influence function for the average Wald estimand, we derive  locally {semiparametric}  efficient  estimators that are \emph{multiply robust} in the sense that they are consistent in the union of three different observed data models. This is in contrast to previous estimators for the LATE  \citep[e.g.][]{tan2006regression,ogburn2015doubly},  the ATE \citep[e.g.][]{okui2012doubly,vansteelandt2015robustness} and a related causal parameter, the effect of treatment on the treated (ETT) \citep{tchetgen2013alternative,liu2015identification}, wherein  researchers only obtain  doubly robust  estimators.  
% In related works, \cite{tchetgen2013alternative} propose  sufficient identifying conditions for the conditional average treatment effect on the treated (ETT) given baseline variables and develop corresponding DR estimators, while \cite{liu2015identification} give necessary and sufficient identifying conditions for the marginal ETT and develop a locally efficient DR estimator.  
Furthermore, we discuss the setting of binary outcomes in detail, for which IV methods are not well developed. Towards that end, we propose variation independent
parameterizations of the likelihood so that the parameter space under our model specification is unconstrained. We also propose
bounded estimators for the ATE that always lie in the parameter space, which is $[-1,1]$   for a binary outcome. Throughout we assume both the instrument and the treatment to be binary.

The rest of this article is organized as follows. In Section \ref{sec:framework}, we introduce the IV  set-up, provide various potential outcomes definitions we shall refer to throughout the paper and discuss existing  results on partial identification.  We provide formal identification conditions and corresponding estimation approaches in Sections \ref{sec:identification} and \ref{sec:estimation}, respectively.   In Section \ref{sec:simulation}, we evaluate the finite sample performance of the proposed estimators via simulations. In Section \ref{sec:real_data}, we apply the proposed methods to estimate the causal effect of education on earnings using data from the National Longitudinal Study of Young Men. We end with a brief discussion in Section \ref{sec:discussion}. 

The {\tt R} programs that were used to analyse the data can be obtained from \url{https://dataverse.harvard.edu/dataset.xhtml?persistentId=doi:10.7910/DVN/BIJWCU}.

\section{Framework and notation}
\label{sec:framework}

Consider an observational study with a single follow-up visit. 
Suppose we are interested in estimating the effect of a binary exposure $D$ on an outcome variable $Y$, but the effect of $D$ on $Y$ is subject to confounding by observed variables $X$ as well as  unobserved variables $U$.
A variable Z is called an IV if it satisfies the following assumptions \citep[e.g.][]{didelez2007mendelian}:
\begin{itemize}
	\item[A1] Exclusion restriction: $Z \ind Y \mid (D, U, X)$;
	\item[A2] Independence: $Z \ind U \mid X$;
	\item[A3] Instrumental variable relevance: $Z \nind D\mid X.$
\end{itemize}  Figure \ref{DAG:iv_model} gives causal graph representations \citep{pearl2009causality,richardson2013single} of the
(conditional) IV model.

\begin{figure}[!htbp]
\parbox{.5\textwidth}{
	\vspace{0.9cm}
	\centering
	\begin{tikzpicture}[->,>=stealth',node distance=1cm, auto,]
	%nodes
	\node[est] (Z) {$Z$};
	\node[est, right = of Z] (D) {$D$};
	\node[est, right = of D] (Y) {$Y$};
	\node[shade, below = of D] (U) {$U$};
	\node[est, above = of D] (V) {$X$};
	\path[pil] (Z) edgenode {} (D);
	\path[pil] (D) edgenode {} (Y);
	\path[pil] (U) edgenode {} (D);
	\path[pil] (U) edgenode {} (Y);
	\path[pil] (V) edgenode {} (Z);
	\path[pil] (V) edgenode {} (D);
	\path[pil] (V) edgenode {} (Y);
	\path[double] (Z) edge [bend right] node  {} (D);
	\end{tikzpicture}
	\quad \\ \bigskip (a). A DAG with a bi-directed arrow.
}
\parbox{.5\textwidth}{
		\centering	
	\begin{tikzpicture}[>=stealth, ultra thick, node distance=2cm,
	% \tikzstyle{format} = [draw, ultra thick, minimum size=6mm
	pre/.style={->,>=stealth,ultra thick,black,line width = 1.5pt}]
	\begin{scope}
	\node[name=Z,shape=ellipse splitb, ellipse splitb/colorleft=black, ellipse splitb/colorright=red, 
	ellipse splitb/innerlinewidthright = 0pt,  %Remove the inner 'line'
	/tikz/ellipse splitb/linewidthright = 1pt,   %Make Right line same width as left
	ellipse splitb/gap=3pt, style={draw},rotate=90] {
		{\rotatebox{-90}{$Z$\;}}
		\nodepart{lower}
		{\rotatebox{-90}{\;\textcolor{red}{$z$}}}
	};
	\node[name=D,shape=ellipse splitb, below right=0.65cm and 1.8cm of Z,  ellipse splitb/colorleft=black, ellipse splitb/colorright=red, 
	ellipse splitb/innerlinewidthright = 0pt,  %Remove the inner 'line'
	/tikz/ellipse splitb/linewidthright = 1pt,   %Make Right line same width as left
	ellipse splitb/gap=3pt, style={draw},rotate=90] {
		{\rotatebox{-90}{$D(\tred{z})$\;}}
		\nodepart{lower}
		{\rotatebox{-90}{\;\textcolor{red}{$d$}}}
	};
	\node[thick, name=Y,shape=ellipse,style={draw}, below right =0.05cm and 1.8cm  of D,outer sep=0pt, text width = 8mm]
	{$Y({\color{red}{d}})$
	};
	\draw[pil,->] (Z) to (D);
	\node[est, above right=1.5cm and 0.3cm of D] (V) {$X$};
	\path[pil] (Z) edgenode {} (D);
	\path[pil] (D) edgenode {} (Y);
	\draw[pil, ->] (2.7,2)  to[bend right]  (-0.5,0.4);
	\draw[pil, ->] (3.1,1.7)  --  (3.1,0.4);
	\draw[pil, <->] (3.1,-0.5)  to[bend left]  (-0.5,-0.4);
	\path[pil] (V) edgenode {} (Y);
	\node[shade, below = of D] (U) {$U$};
	\path[pil] (U) edgenode {} (3.3,-0.5);
	\path[pil] (U) edgenode {} (Y);
	%\node[below right = 8mm of anew2]{(a)};
	\end{scope}
	\end{tikzpicture}
	\quad \\ \bigskip	(b). A SWIG with a bi-directed arrow.
}
	\caption{Causal graphs representing an instrumental variable model. The bi-directed arrow between $Z$ and $D$ indicates potential unmeasured common causes of $Z$ and $D$. Variables $X,Z,D,Y$ are observed; $U$ is unobserved. The left panel gives a causal Directed Acyclic Graph \citep{pearl2009causality} with a bi-directed arrow, and the right panel gives a Single World Intervention Graph \citep{richardson2013single} with a bi-directed arrow.}
	\label{DAG:iv_model}
\end{figure}

An alternative definition of IV is based on the potential outcome framework \citep{neyman1923applications,rubin1974estimating}, under which we assume $D(z)$, the potential exposure if the instrument would take value $z$ to be well-defined \citep[the Stable Unit Treatment Value Assumption,][]{rubin1980comment}. Similarly, we assume $Y(d,z)$, the response that would be observed if a unit were exposed to $d$ and the instrument had taken value $z$ to be well-defined. The definition of IV under this framework can be derived from a single-world intervention graph (SWIG) \citep{richardson2013single}  similar to the one in Figure \ref{DAG:iv_model}(b), but without the bi-directed arrow $Z\leftrightarrow D(z)$. In particular, A1 is replaced with the assumption A1$^\prime$ that $\forall z, Y(z,d) = Y(d)$ and A2 is replaced with the assumption A2$^\prime$ that $\forall z, d, Z \ind (D(z), Y(d)) \mid X. $ We refer interested readers to \cite{dawid2003causal} and \cite{richardson2014ace} for discussions on the connections and differences between these two definitions.

In this article, we assume A1$^\prime$, A2, A3 and the additional assumption A4 that  $Y(d) \ind (D,Z) \mid (X,U).$ The last assumption may also be read (via d-separation) from the SWIG in Figure \ref{DAG:iv_model}(b).
Note we allow for unmeasured common causes of $Z$ and $D$, so that the instrument $Z$ and exposure $D$ may be associated simply because they share a latent common cause.  { This is important as in observational settings,
it	may be difficult to ensure that one has measured all common causes of $Z$ and $D$.} However, if $Z$ represents a randomized experiment, then one might be willing to make the stronger assumption that the causal effect of Z on D is  unconfounded so that $Z \ind D(z)\mid X$.  This would allow us to remove the bi-directed arrow from the graph in Figure \ref{DAG:iv_model}. \cite{hernan2006instruments} refer to such IVs as ``causal'' IVs. In this case, under the principal stratum framework \citep{frangakis2002principal},  the population can be divided into four strata based on values of $(D(1), D(0))$ as in Table \ref{table:defG}.
\begin{table}[!htbp]
	\centering
	\caption{Principal stratum describing potential complier status $(D(1),D(0))$  }\label{table:defG}
	\begin{tabular}{cc|cc}
		\toprule
		$D(1)$ & $D(0)$ & Principal stratum & Abbreviation \\
		\midrule
		1 & 1 &  Always taker & AT \\
		1 & 0 &  Complier  & CO \\
		0 & 1 &  Defier & DE \\
		0 & 0 &  Never taker & NT \\
		\bottomrule
	\end{tabular}
\end{table}
Under a further monotonicity assumption that $P(D(1) \geq D(0)) = 1,$ one is able to identify the local average treatment effect $LATE = E[Y(1)-Y(0)\mid D(1)> D(0)]$ \citep{imbens1994identification,abadie2003semiparametric}. However, even with a causal IV and the monotonicity assumption, the average treatment effect is not identifiable. The fundamental difficulty is that without further assumptions,
it is not possible to identify the treatment effect in stratum AT or NT  as the subjects in these strata always (or never)  take the treatment. 
 Instead, in the case where $Z, D, Y$ are binary and $X$ is an empty set, \cite{richardson2014ace} derive sharp bounds for the ATE under our IV assumptions. See also  
  \cite{balke1997bounds} and \cite{chesher2010instrumental} for sharp bounds under other definitions of  IV.

%An alternative definition of instrumental variable involves  use of potential outcomes \citep{angrist1996identification}. We refer interested readers to 

%\subsubsection*{IV assumptions}
%\begin{itemize}
%	\item[A.1]  Independence: $Z \ind U \mid X$ or $Z \ind Y(z,d),D(z) \mid X.$
%	\item[A.2] Exclusion restriction: $Z \ind Y \mid X,U$  or $Y(z,x) = Y(x)$
%	\item[A.3] Relevance: $Z \nind D \mid X$ 
%	\item [A.3*]  Causal relevance: $E[D(z)]$ is not a constant of $z$
%	\item [A.4] Monotonicity: $D(1) \geq D(0), a.s.$
%\end{itemize}
%Several remarks
%\begin{itemize}
%	\item The DAG version can be found in \cite{didelez2007mendelian}, and the potential outcome version can be found in \cite{imbens1994identification} and \cite{angrist1996identification}.
%	\item Note that A.3* is stronger than A.3. \cite{hernan2006instruments} discuss their difference in detail. 
%	\item \citet[][Proposition A.1]{tan2006regression} shows that A.1 (in the form of $Z \ind (Y(1),Y(0))\mid X$) and A.4 are equivalent to the structural model that 
%	\begin{flalign*}
%	D(z) &= 1\{\pi(z, X) \geq U\} \\
%	Z &\ind (Y(0), Y(1)) \mid X, \quad Z \ind U \mid X,
%	\end{flalign*}
%	where $pi(z, x) = P(D=1\mid Z=z, X=x), U \sim U(0,1)$.
%\end{itemize}
%

\section{Identification of the average treatment effect}
\label{sec:identification}

In this section, we consider the identification problem for the average treatment effect 
$$
		ATE = E[Y(1) - Y(0)]. $$ Specifically,
in the following Theorem \ref{thm:identification}, we give two general no-interaction assumptions for identification of the ATE. The proof is left to the Appendix. 

\begin{theorem}
	\label{thm:identification}
%	Assume that $E[D\mid Z=1, X] - E[D\mid Z=0, X] \neq 0, a.e.$.
	Under our IV model that assumes A1$^\prime$, A2, A3 and A4, the ATE is identifiable if \emph{either of} the following assumptions holds:
	\begin{enumerate}
		\item[A5.a] There is no additive $U-Z$  interaction in  $E[D\mid Z, X,U]$:
		$$
				E[D\mid Z=1,X,U] - E[D\mid Z=0,X,U]  = E[D\mid Z=1,X] - E[D\mid Z=0, X];
		$$
		\item[A5.b] There is no additive $U-d$ interaction in 	$E[Y(d) \mid X,U]$:
				$$
				E[Y(1) - Y(0)\mid X,U] = E[Y(1) - Y(0) \mid X].
				$$
	\end{enumerate}

	Furthermore,  under either of these assumptions,  the conditional ATE equals the conditional Wald estimand
	$$
			ATE(x) = E[Y(1) - Y(0)\mid X=x] = \dfrac{E[Y\mid Z=1,X=x] - E[Y\mid Z=0,X=x]}{E[D\mid Z=1,X=x] - E[D\mid Z=0,X=x]} \triangleq \dfrac{\delta^Y(x)}{\delta^D(x)}  \triangleq \delta(x)
	$$
	and the marginal ATE equals the average Wald estimand $ATE = E_X \delta(X) \triangleq \Delta.$
\end{theorem}

Throughout 	for notation convenience, we say A5 holds if either A5.a or A5.b holds.
A5.a is a generalization of the stage-I model \eqref{eqn:lsem1}.
It states that upon conditioning on measured covariates, 
%there is no unmeasured effect modifiers for the instrumental effect on the exposure. 
no \emph{unmeasured} confounder of the $D-Y$ association interacts with the IV on the additive scale in predicting the exposure. With a causal IV such that $Z\ind D(z) \mid (X, U)$ and $Z\ind D(z) \mid X,$ A5.a can be written in a similar form as A5.b: $E[D(1)-D(0)\mid X,U] = E[D(1)-D(0)\mid X]$. 
%We also note that  due to the IV independence assumption,   $E[D(z)\mid X,U] = E[D\mid Z,X,U]$ so that   
On the other hand, A5.b states that conditioning on measured covariates, 
%there is no unmeasured effect modifiers for the exposure effect on the outcome. 
no \emph{unmeasured} confounder of the $D-Y$ association modifies the causal effect of $D$ on the mean of $Y$ on the additive scale. Similar to \cite{hernan2006instruments}'s treatment homogeneity assumption, A5.b has the attractive property that it is guaranteed to hold under the null hypothesis of no causal effect for all units.   Assumptions A5 has an important implication for the \emph{design} of observational studies: even if a randomized instrument is available, it is still important to collect as many causes of the exposure and outcome as possible in the hope that no residual effect modification remains within strata defined by covariates $X$.

%{\bf Eric: the story on sufficient cause? (c.f. Tyler's work)}

It is interesting to note that with a causal IV, $\delta(X)$ may also be interpreted as $LATE(X) = E[Y(1) - Y(0)\mid D(1) = 1, D(0)=0, X]$ under the monotonicity assumption that $D(1) \geq D(0), a.e.$ \citep{imbens1994identification}. Hence with a causal IV, the monotonicity assumption and A5 together imply  the latent ignorability assumption $LATE(X) = ATE(X)$ \citep{frangakis1999addressing,angrist2013extrapolate,aronow2013beyond}.  Similarly, under a no-current-treatment-value-interaction assumption, $\delta(X)$ can also be interpreted as $ETT(X) = E[Y(1)-Y(0)\mid D=1, X]$  \citep{hernan2006instruments}. Consequently the no-current-treatment-value-interaction assumption and A5 together imply that  $ETT(X) = ATE(X).$

However, even under the monotonicity assumption or the no-current-treatment-value-interaction assumption, our assumption does \emph{not} imply that the marginal ATE is the same as the marginal LATE or the marginal ETT.  The latter assumption is questionable as the complier and the treated arm may both be highly selective groups of the population.  To see their differences, note that under different sets of identification assumptions described above,
\begin{flalign*}
ATE &= E_X ATE(X) =  E_X \dfrac{\delta^Y(X)}{\delta^D(X)}; \\
LATE &= E_{X\mid D(1)>D(0)} LATE(X) = \dfrac{E_X \delta^Y(X)}{E_X \delta^D(X)}; \numberthis \label{eqn:late}\\
ETT &= E_{X\mid D=1} ETT(X) = E_{X\mid D=1} \dfrac{\delta^Y(X)}{\delta^D(X)},
\end{flalign*}
where the second equality in \eqref{eqn:late} is due to \citet[][Theorem 3.1]{abadie2003semiparametric}. One can also see from \eqref{eqn:late} that with a causal IV, in the case where A5.a and A5.b are incorrect but the monotonicity assumption  is correct, our estimand $E_X \delta(X)$ can still be interpreted as the LATE for a complier population whose covariate distribution matches that of the full study population \citep{aronow2013beyond}; similarly for the case where  only the no-current-treatment-value-interaction assumption is correct. 
%In the Appendix we show

\begin{remark}
	When neither A5.a or A5.b holds, in general the Wald estimand differs from $ATE$. In this case, using a similar argument as \cite{vanderweele2008sign}'s, one may show that under additional assumptions, it is still possible to determine the sign of the bias for estimating the ATE using an unbiased estimator of the Wald estimand.
%	 when monotonicity relationships hold between the unmeasured confounding variables $U$ and the instrumental effect on the exposure, and between the unmeasured confounding variables $U$ and the exposure  effect on the outcome. 
We defer the detailed discussion to Proposition  \ref{prop:wald} in Appendix \ref{appendix:wald}.
\end{remark}

\section{Bounded, efficient and multiply robust estimation}
\label{sec:estimation}

In this section we describe estimation methods for the average Wald estimand
$E_X\delta(X) \triangleq \Delta$. In principle, one can estimate $\Delta$ by first estimating $\delta(x)$ and then taking expectation with respect to the empirical distribution of $X$.  The quantity $\delta(x)$ has been the inferential target of  previous papers since under the identification assumptions discussed above,  it can be interpreted as the conditional LATE \citep{imbens1994identification} or the conditional ETT \citep{hernan2006instruments}.
However, with the exception of g-estimators, existing estimators for $\delta(x)$ are not guaranteed to be \emph{bounded} \citep{robins2007comment} in the sense that they do not necessarily fall within the parameter space of $E[Y(1) - Y(0)\mid X=x]$. This is particularly relevant when the outcome $Y$ is binary, in which case the parameter space for both $\delta(x)$ and $\Delta$ is  $[-1,1].$

In what follows we introduce three classes of estimators for $\Delta$. We also propose {bounded} versions of these estimators which are guaranteed to fall within the parameter space with binary $Y$. In addition, we  show that these estimators are consistent and asymptotically normal (CAN) under different sets of model assumptions:
	\begin{enumerate}
		\item[$\m_1$:] models for $\delta \left( X\right) ,\delta^D(X) $, $p_{0}^Y
		\left( X\right) \equiv E[Y\mid Z=0, X] $ and $p_{0}^{D}(X) \equiv E[D\mid Z=0, X]$ are correct;
		\item[$\m_2$:]  models for $\delta^D(X)$ and the conditional density of $Z$ given $X$, denoted as $f(Z\mid X)$    are correct;
		\item[$\m_3$:]  models for $\delta(X)$ and $f(Z\mid X)$   are correct.
	\end{enumerate}
We then propose estimators that are \emph{multiply robust} in the sense that they are CAN if one, but not necessarily more than one of models $\m_1$, $\m_2$, $\m_3$ is correct. Moreover, they are \emph{locally efficient}  in that they achieve the semiparametric efficiency bound for the union of $\m_1,\m_2,\m_3$, denoted as $\m_{\text{\it union}}$, at the intersection of  $\m_1$, $\m_2$, $\m_3$.  

We note that in general, the identification assumptions A1$^\prime$ -- A5 imply certain constraints on the observed data law, known as instrumental inequalities \citep{pearl1995testability}. In Proposition \ref{prop:constraints} we discuss these constraints for the binary IV model, in which case the instrumental inequalities are known to be sharp \citep{bonet2001instrumentality}. The proof is left to the on-line supplementary materials.
\begin{proposition}
	\label{prop:constraints}
	\quad
	\begin{itemize}
		\item[(i)] If $Z,D,Y$ are binary, then the canonical IV assumptions A1$^\prime$, A2, A3 and A4 impose testable implications on the law of $(Z, D, Y, X)$:
		\begin{equation}
		\label{eqn:iv_ineq}
			P(Y=y, D=d\mid Z=1,X) + P(Y=1-y, D=d\mid Z=0,X) \leq 1, \quad  y =0,1, d=0,1.
		\end{equation}
		However, they do not have testable implications on the laws modeled in $\m_1, \m_2, \m_3$, which are $P(Y=1\mid Z, X), P(D=1\mid Z, X)$ and $f(Z\mid X).$
		\item[(ii)] The no-interaction assumption A5 does not have testable implications on the law of observed data $(Z,D,Y,X)$.
	\end{itemize}
\end{proposition}

% Despite of this, previous estimation methods developed for $\delta(X)$ may be instrumental for our goal of estimating $\Delta$.  We now first introduce three different estimators of $\Delta$ that draw inspiration from existing estimators but improve on them so that they are suitable even when the outcome is binary. We then introduce a multiply robust estimator that combines this three estimators. 
 
% Previous regression and weighting estimators of $\delta(X)$, however, are not directly applicable  to our case with binary $Y$. We hence discuss them separately in  in Section \ref{sec:regression_estimator} and \ref{sec:ipw}, respectively. 
% 
% Finally in Section \ref{sec:multiply_robust} we 
% construct a locally semiparametric efficient estimator that is multiply robust in the sense that it is consistent and asymptotically normal (CAN) in the union of the three parametric models. In other words, our estimator is CAN if one, but not necessarily more than one of the ... is true. 

\subsection{Regression-based estimators}
\label{sec:regression_estimator}

We first discuss regression-based estimation  of $\Delta$. Some existing proposals for estimating $\delta(x)$ and hence $\Delta$ have relied on separate regression-based estimators for  $\delta^Y(x)$ and $\delta^D(x)$.  For example, \cite{frolich2007nonparametric} imposes models on $E[Y\mid Z=z, X]$ and $E[D\mid Z=z, X]$, while \cite{tan2006regression} places models on $E[Y\mid Z=z, D=d, X]$ and $E[D\mid Z=z, X]$. 
 These models, although intuitive, may produce  estimators of $\delta(x)$ that are not bounded  with $Y$ binary.  Although one can choose suitable models for $E[Y\mid Z=z,X]$ (or $E[Y\mid Z=z,D=d,X]$) and $E[D\mid Z=z,X]$  to constrain  $\delta^Y(X)$ and $\delta^D(X)$ within $[-1,1]$, generally there is no guarantee that $\delta(x) = \delta^Y(x) / \delta^D(x)$ and hence $\Delta = E_X \delta(X)$ lie between -1 and 1.

To remedy this difficulty in the case of binary $Y$, we instead impose models on $\delta(X)$ directly, such as
\begin{equation}
\delta(X; \alpha) = \tanh(\alpha^T X) = \dfrac{e^{2 \alpha^T X}-1}{e^{2\alpha^T X}+1}, \numberthis\label{eqn:delta}
\end{equation}
which guarantees that $\delta(X) \in [-1,1].$  Nuisance models are then needed to  allow for maximum likelihood estimation of $\alpha$. Prior to our work, \cite{okui2012doubly} and \cite{vansteelandt2015robustness}  choose the nuisance model to be   $E[Y-\delta(X) D\mid  X; \beta_{\text{okui}}]$. However, with a binary $Y$, a model such as \eqref{eqn:delta} is variation dependent with  $E[Y-\delta(X)D \mid X; \beta_{\text{okui}}]$, as with an arbitrary choice of $(\alpha, \beta_{\text{okui}}),$ $E[Y\mid  X; \alpha, \beta_{\text{okui}}] = E[Y-\delta(X)D \mid X; \beta_{\text{okui}}] + \delta(X; \alpha) E[D\mid X]$ may not lie between 0 and 1.  Consequently, the parameter space of $(\alpha, \beta_{\text{okui}})$ is a constrained space in $\mathbb{R}^{2p},$  making maximum likelihood estimation and asymptotic analysis difficult; here $p$ refers to the dimension of $X$.    Instead, following \cite{richardson2017modeling},  our choice of nuisance models is 
$$(\delta^D(X; \beta), OP^D(X; \zeta), OP^Y(X; \eta)),$$ 
where $OP^D(X) =[ p_1^D(X) p_0^D(X)]/ [(1-p_1^D(X))(1-p_0^D(X))], OP^Y(X) =[ p_1^Y(X) p_0^Y(X)]/ [(1-p_1^Y(X))(1-p_0^Y(X))], p_1^D(X) \equiv P(D=1\mid Z=1,X), p_1^Y(X) \equiv P(Y=1\mid Z=1,X),$ and $\delta^D(X;\beta) = \tanh(\beta^T X) \in [-1,1].$  Proposition \ref{prop:one-to-one} shows that our models  provide a variation independent parameterization of the likelihood $(P(Y=1\mid Z,X), P(D=1\mid Z,X))$ so that the parameter space of $(\alpha, \beta, \zeta, \eta)$ is unconstrained.

\begin{proposition}
	\label{prop:one-to-one}
	For any realization of $X$, the mapping given by
	\begin{equation}
	\label{eqn:mapping}
	(\delta(X), \delta^D(X), OP^D(X), OP^Y(X))  \rightarrow (p_0^D(X),p_1^D(X), p_0^Y(X), p_1^Y(X))
	\end{equation}  is a diffeomorphism between the interiors of their domains, which are $(-1,1)^2 \times (\mathbb{R}^+)^2$ and $(0,1)^4$, respectively, where $\mathbb{R}^+ = (0,\infty)$. Moreover, it is available in closed form:
	\begin{flalign*}
	p_0^D(X) &= \dfrac{1}{2(OP^D(X)-1)} \left\{ 
	OP^D(X)(2-\delta^D(X))+\delta^D(X)-  \textcolor{white}{\sqrt{OP^D(X)}} \right. \numberthis \label{eqn:p0d}   \\
	&\left.		\sqrt{\{OP^D(X)(\delta^D(X)-2)-\delta^D(X)\}^2+4OP^D(X)(1-\delta^D(X))(1-OP^D(X))}\right\}; \\[2pt]
	p_1^D(X) &= p_0^D(X) + \delta^D(X); \\[2pt]
	p_0^Y(X) &= \dfrac{1}{2(OP^Y(X)-1)} \left\{ 
	OP^Y(X)(2-\delta(X)\delta^D(X))+\delta(X)\delta^D(X)-  \textcolor{white}{\sqrt{OP^Y(X)}} \right. \numberthis \label{eqn:p0y}   \\
	&\left.		\sqrt{\{OP^Y(X)(\delta(X)\delta^D(X)-2)-\delta(X)\delta^D(X)\}^2+4OP^Y(X)(1-\delta(X)\delta^D(X))(1-OP^Y(X))}\right\}; \\[2pt]
	p_1^Y(X) &= p_0^Y(X) + \delta(X) \delta^D(X).\end{flalign*}
\end{proposition}

In principle, any choice of nuisance functions that make the mapping \eqref{eqn:mapping} a diffeomorphism would suffice. We choose odds products as they are simple and the mapping \eqref{eqn:mapping} is available in closed \hbox{form}.  Under this parameterization, in model $\m_1$  we say $p_0^D(X; \beta, \eta)$ is correctly specified if the models $\delta^D(X; \beta)$ and $OP^D(X; \eta)$ are correct, and  $p_0^Y(X; \alpha, \beta, \zeta)$ is correctly specified if the models  $\delta(X;\alpha), \delta^D(X; \beta)$ and $OP^Y(X;\zeta)$ are correct.

%\begin{equation*}
%\delta^D(X; \beta) = \tanh(\beta^T X) = \dfrac{e^{2 \beta^T X}-1}{e^{2\beta^T X}+1}  
%%\numberthis\label{eqn:delta_D}
%\end{equation*}
%which guarantees that $\delta^D(X) \in [-1,1].$ 

%Note  that $\delta^Y(X; \alpha, \beta) = \delta(X; \alpha) \delta^D(X;\beta) $ also falls within $[-1,1]$. Moreover, instead of 
% imposing models on $p_0^Y(X)$ and $p_0^D(X)$ that are \emph{variation dependent} of models on $\delta^D(X)$ and $\delta(X)$, we follow \cite{richardson2017modeling} to  assume models  
% These models are chosen because they are both   variation independent of models for $\delta^D(X)$ and $\delta(X)$ and relatively simple. Moreover, the mapping from 

% In particular the right hand sides of the equations in Proposition \ref{prop:one-to-one} are guaranteed to fall between 0 and 1.

%The four models $\delta(X; \alpha), \delta^D(X; \beta), \log(OP^D(X; \eta))$ and $\log(OP^Y(X;\zeta))$ are variation independent \citep{richardson2017modeling} so that the parameter space of $(\alpha, \beta, \eta, \zeta)$ is unconstrained. 

Two-step (unconstrained) maximum likelihood can then be used for estimation of these parameters. Specifically, 
let ($\hat\beta_{\text{\it 2mle}}$,$\hat\eta_{\text{\it 2mle}}$) denote the solution to the score equations corresponding to the likelihood of $D$ conditional on $Z$ and $X$: $\mathbb{P}_n S(D\mid Z,X; \beta, \eta) =0,$ and ($\hat\alpha_{\text{\it 2mle}}, \hat\zeta_{\text{\it 2mle}}$) denote the solution to the score equations corresponding to the likelihood of $Y$ conditional on $Z,X$ and $\hat{\beta}_{\text{\it 2mle}}$:  $\mathbb{P}_n S(Y\mid Z,X; \alpha, \hat{\beta}_{\text{\it 2mle}}, \zeta) =0.$ 
The bounded regression-based estimator of $\Delta$ is given as 
$$
		\hat{\Delta}_{\text{\it b-reg}}  = \mathbb{P}_n \delta(X; \hat{\alpha}_{\text{\it 2mle}}),
$$
where $\mathbb{P}_n$ denotes the empirical average: $\mathbb{P}_n X =\dfrac{1}{n} \sum\limits_{i=1}^n X_i$.  
Theorem \ref{thm:reg} summarizes the key properties  of $\hat{\Delta}_{\text{\it b-reg}}.$ The proof is left to the on-line supplementary materials.
\begin{theorem}
	\label{thm:reg}
Under standard regularity conditions,	$\hat{\Delta}_{\text{\it b-reg}}$ is CAN in model $\m_1$. 
\end{theorem}

	In some settings researchers may be willing to assume a causal IV and make the monotonicity assumption, under which $\delta^D(X)$ lies between 0 and 1. To respect this range, one may instead fit a logistic model  $\delta^D(X;\beta) = \text{expit}(\beta^T X)$ and one can show that the mapping \eqref{eqn:mapping} is still a diffeomorphism between their domains. 
	
%	 In this case, $\delta^Y(X;\alpha,\beta) = \delta(X;\alpha) \delta^D(X;\beta)$ still falls within $[-1,1]$ as long as $\delta(X;\alpha)\in [-1,1].$

%This parameterization has the advantage that there is a one-to-one mapping between $(\text{arctanh}(\delta^D(X)), \log(OP^D(X)))$ and $(E[Y\mid Z=1,X], E[Y\mid Z=0,X])$; at the same time, the domain of $(\text{arctanh}(\delta^D(X)), \log(OP^D(X)))$ is $\mathbb{R}^2$ so that there are no restrictions on the models that can be placed on $\text{arctanh}(\delta^D(X))$ and $\log(OP^D(X)))$.
%
%
% it is well known that $\delta^D(X)$ and $p^D_0(X)$ are \emph{variation dependent} in the sense that the range of $p^D_0(X)$ depends on the value of $\delta^E(X)$ \citep{richardson2017modeling}; similarly with $\delta^Y(X) = \delta(X)\delta^D(X) $ and $p^Y_0(X)$. We hence follow \cite{richardson2017modeling} to place modeling restrictions on $\delta(X), \delta^D(X), 

%
%\subsection{g-estimation}
%
%As shown by \cite{hernan2006instruments}, under the structural mean model that 
%$$
%		E[Y(1) - Y(0)\mid D=1,Z=z,X] = \psi_0^{\ast}(X;\alpha),
%$$
%we have $\delta(X;\alpha) = \psi_0^{\ast}(X;\alpha)$. Therefore, estimation of $\alpha$ and hence $\Delta$ can be based on g-estimation methods \citep{robins1994correcting} that solves the following estimation equations:
%\begin{equation}
%\label{eqn:g_est}
%E\left\{ h(X)(Y-D \delta(X;\alpha)) (Z-\pi(X))  \right\} = 0,
%\end{equation}
%where $\pi(X) = P(Z=1\mid X)$ and $h(X)$ is an arbitrary function with the right dimension. 

\subsection{Inverse probability weighting estimation}
\label{sec:ipw}
In contrast with  regression-based estimation,   inverse probability weighting (IPW)  avoids placing modeling restrictions on the outcome. Instead, it only assumes models in $\m_2$.
%These are in parallel to the propensity score methodologies to deal with measured confounding.
This is considered advantageous as IPW separates the design stage from the analysis stage in the sense that models in $\m_2$ are specified prior to seeing any outcome data, and thus helps prevent selecting models that favor ``publishable'' results \citep{rubin2007design}. 

IPW estimation assumes models $\delta^D(X;\beta)$ and $f(Z\mid X;\gamma).$ Let $\hat{\gamma}_{\text{\it mle}}$ be the maximum likelihood estimator of $\gamma$, and  $\hat{\beta}_{\text{\it ipw}}$ solves the following equation:
\begin{equation}
\label{eqn:beta}\mathbb{P}_n \left[h_1(X) \left( \dfrac{D(2Z-1)}{f(Z\mid X;\hat{\gamma}_{\text{\it mle}})} - \delta^D(X;\beta)  \right)  \right] = 0,
\end{equation}
where $h_1(X)$ is a vector function of the same dimension as $\beta$, such as $h_1(X)=X$. An IPW estimator of $\Delta$ is  defined as follows:
\begin{equation}
\label{eqn:ipw2}
\hat{\Delta}_{\text{\it ipw}} = \mathbb{P}_n  \dfrac{Y}{\delta^D(X; \hat{\beta}_{\text{\it ipw}})}\dfrac{2Z-1}{f(Z\mid X; \hat{\gamma}_{\text{\it mle}})}.    
\end{equation}

One problem with $\hat{\Delta}_{\text{\it ipw}}$ is  that it is not bounded with a binary $Y$.  To remedy this, one may project \eqref{eqn:ipw2} onto a bounded working model. Specifically,  let $\hat\alpha_{working}$  solve the following  equation:
\begin{equation*}
\label{eqn:b-ipw}
	\mathbb{P}_n  \left[h_2(X) \left( \dfrac{Y}{\delta^D(X;\hat{\beta}_{\text{\it ipw}})} \dfrac{2Z-1}{f(Z\mid X;\hat{\gamma}_{\text{\it mle}})} - \delta(X;\alpha) \right) \right] = 0,
\end{equation*}
where $\delta(X;\alpha)$ falls within $[-1,1]$ and $h_2(X)$ is a vector function of the same dimension as $\alpha$, such as $h_2(X)=X$. 
A bounded IPW estimator is then defined as
\begin{flalign*}
\hat{\Delta}_{\text{\it b-ipw}}	&= \mathbb{P}_n \delta(X;\hat\alpha_{working}). \numberthis \label{eqn:ee_ipw}  
\end{flalign*}

%Note that $\delta(X;\alpha)$ is just a working model here. Even if it is not correctly specified, \eqref{eqn:b-ipw} and \eqref{eqn:ee_ipw} 
%still implies \eqref{eqn:ipw2} and hence  yields a CAN estimator of $\Delta$ as long as the models  $\delta^D(X;\beta)$ and $f(Z\mid X;\gamma)$ are correctly specified.   On the other hand, even if it is correctly specified,   $\hat\Delta_{\text{\it b-ipw}}$ may not be consistent outside of $\m_2$. 

Theorem \ref{thm:ipw} summarizes the properties of $\hat{\Delta}_{\text{\it ipw}}$ and $\hat{\Delta}_{\text{\it b-ipw}}$. The proof is left to the on-line supplementary materials.

\begin{theorem}
	\label{thm:ipw} 
	Under standard regularity conditions and the positivity assumption that both $\delta^D(X)$ and $f(Z\mid X)$ are bounded away from 0, $\hat{\Delta}_{\text{\it ipw}}$ and $\hat{\Delta}_{\text{\it b-ipw}}$ are CAN in $\m_2$, regardless of whether or not the model $\delta(X;\alpha)$ is correct.
\end{theorem}

\subsection{G-estimation}
\label{sec:g-estimation}

Estimation of $\Delta$ can also be based on g-estimation under model $\m_3$. 
 Specifically, let $\hat{\alpha}_g$ solve the following equation:
\begin{equation}
\label{eqn:g_est}
\mathbb{P}_n \left\{ h_3(X)(Y-D \delta(X;\alpha)) \dfrac{2Z-1}{f(Z\mid X;\hat{\gamma}_{\text{\it mle}})}  \right\} = 0,
\end{equation}
where $f(Z\mid X; \hat{\gamma}_{\text{\it mle}})$ is defined in Section \ref{sec:ipw} and  $h_3(X)$ is a vector function of the same dimension as $\alpha$, such as $h_3(X) = X$.  The g-estimator of $\Delta$ is given as
$$
		\hat{\Delta}_g = \mathbb{P}_n \delta(X;\hat{\alpha}_g).
$$
It is interesting to note that \eqref{eqn:g_est} coincides with the g-estimating equation \citep{robins1994correcting} for estimating parameter $\alpha$ in the following structural mean model for the conditional ETT:
\begin{equation}
\label{eqn:ett}
E[Y(1) - Y(0)\mid D=1,Z,X] = \psi_0^{\ast}(X;\alpha),
\end{equation}
replacing $\psi_0^{\ast}(X;\alpha)$ with   $\delta(X;\alpha)$.
This is because \eqref{eqn:ett} implies the  no-current-treatment-value-interaction assumption, 
so that   $\psi_0^{\ast}(X;\alpha) = \delta(X;\alpha)$.

 Theorem \ref{thm:g} summarizes the properties of $\hat{\Delta}_g$. The proof is very similar to that of Theorem \ref{thm:reg} and hence omitted. 
 \begin{theorem}
 	\label{thm:g}
 	Under standard regularity conditions and the positivity assumption that $f(Z\mid X)$ is bounded away from 0,	$\hat{\Delta}_{\text{\it g}}$ is CAN in model $\m_3$. 
 \end{theorem}

%In practice, one may also specify a model for $\pi(X)$, such as
%\begin{equation}
%\label{eqn:prop}
%\pi(X;\gamma) = \text{expit}(\gamma^T X).
%\end{equation}
%Estimation of $\Delta$ can then be based on averaging over the empirical distribution of $X$:
%$$
%\hat{\Delta}_{\text{\it g}} = \mathbb{P}_n \delta(x_i, \hat{\alpha}_g),
%$$
%where $\alpha_g$ is obtained from solving estimating equations \eqref{eqn:g_est} and \eqref{eqn:prop}. Under regularity conditions, it is easy to show that $\hat{\Delta}_{\text{\it g}}$ is  consistent and asymptotically normal (CAN) under correct specification of models in $\m_2$. 

G-estimation provides a plug-in estimator.  When $Y$ is binary, to ensure that $\hat{\Delta}_g$ lies between $-1$ and $1$, one only needs to choose an appropriate model for $\delta(X)$ that respects the model constraints, such as model \eqref{eqn:delta}.

\subsection{Multiply robust estimation}
\label{sec:multiply_robust}

We have so far described three classes of estimators that are CAN in three different models $\m_1$, $\m_2$, $\m_3$.
Because when 
$X$ is sufficiently high dimensional, one cannot be confident that any of these models is correctly
specified, it is of interest to develop a multiply robust estimation approach, which is guaranteed to
deliver valid inferences about $\Delta$ provided that one, but not necessarily more than one of models $\m_1$, $\m_2$, $\m_3$ is  correctly specified. That is, we aim to construct an estimator that is CAN in the union model $\m_{\text{\it union}}$. 
The following theorem provides the basis for our estimator.
\begin{theorem}
	\label{thm:efficient_score}
	The efficient influence function for $\Delta$ in the union model $\Mu$ is given by
	\begin{equation*}
	\label{eqn:eif}
	EIF_\Delta =  \dfrac{2Z-1 }{f(Z\mid X) } \dfrac{1}{\delta^D(X)}  \left( {Y}- D\delta(X) - p_0^Y(X) + p_{0}^{D}(X)  \delta(X) \right) +\delta \left( X\right) -\Delta.
	\end{equation*} 
	This  coincides with the efficient influence function for $\Delta$ in the nonparametric model $\m_{\text{\it non}}$, in which no restrictions are placed on the distribution of observed data $(Y,D,Z,X).$
\end{theorem}

%\begin{remark}
%	One can show that the efficient influence function in the union model $\Mu$ is also $EIF_\Delta$.  Consequently, the nonparametric variance bound is the same as the semiparametric variance bound in $\Mu$. See \cite{robins2001comment} for related results in a general context.
%\end{remark}

We now construct a locally efficient estimator based on $EIF_\Delta$ and show that it is CAN in $\Mu.$ We first discuss the case of continuous outcomes, for which the range of $Y$ is unrestricted.  Our estimator requires estimation of parameters in  models $\delta(X;\alpha),$ $\delta^D(X;\beta),$ $f(Z\mid X; \gamma),$ $p_0^D(X; \theta)$ and $p_0^Y(X;\iota)$, where $\theta$ and $\iota$ are parameters indexing models for $p_0^D(X)$ and $p_0^Y(X)$, respectively. 
%It follows that the solution $\hat{\Delta}_{\text{\it eff}}$ to 
%\begin{equation}
%\label{eqn:ee_unknown_general}
%	\mathbb{P}_n EIF_\Delta = 0
%\end{equation} 
%has asymptotic variance equal to the nonparametric variance bound for estimating $\Delta$.  
%$\hat{\Delta}_{\text{\it eff}}$ is also multiply robust such that  $E[EIF_\Delta] = 0$ given prior knowledge of the \emph{true functions} in one of the models $\m_1$, $\m_2$, $\m_3$. 
%
%Of course $\hat{\alpha}_{\text{\it eff}}$ is not feasible as it depends on unknown quantities 
%%We propose to use $EIF_\Delta$ as an estimating equation for $\Delta$ which under certain conditions can be made multiply robust wrt to required required model for unknown nuisance 
%$\delta(X), \delta^D(X), f(Z\mid X),   p_0^D(X)$ and $p_0^Y(X)$. When $Y$ is unconstrained (as is in the case of continuous $Y$) and
%$X$ is high dimensional or has several continuous elements, common practice is to specify
%nuisance parametric models for these population quantities: 
To show that our estimator is multiply robust, it must be established that consistent estimators of model parameters  can be obtained under each of $\m_1$, $\m_2$, $\m_3$ without further model assumptions. Although $\gamma, \theta$ and $\iota$ can be estimated based on maximum likelihood, estimation of $\alpha$ or $\beta$ relies on additional nuisance models as the model $\delta(X;\alpha)$ or $\delta^D(X;\beta)$ 
does not give rise to any partial likelihood by itself. Multiply robust estimation requires construction of a consistent estimator of $\beta$ in the union of $\m_1$ and $\m_2$, and  likewise,  a consistent estimator of $\alpha$ in the union of $\m_1$ and $\m_3$. 

We achieve these goals using doubly robust g-estimation  \citep{robins1994correcting}.  Specifically, let $\hat{\beta}_{\text{\it dr}}$ solve
\begin{equation}
\label{eqn:ee_beta}
\mathbb{P}_n h(X)\left\{ D-\delta^D(X;\beta) Z-p_{0}^{D}(X; \hat{\theta}_{\text{\it mle}}) \right\} \dfrac{
	2Z-1} {f(Z\mid X; \hat{\gamma}_{\text{\it mle}})} = 0
\end{equation}
and $\hat{\alpha}_{\text{\it dr}}$ solve
\begin{equation}
\label{eqn:ee_delta}
\mathbb{P}_n g(X)\left\{ Y- D\delta \left( X; \alpha \right) -p_{0}^Y \left( X; \hat{\iota}_{\text{\it mle}} \right) +p_{0}^{D}(X; \hat{\theta}_{\text{\it mle}})\delta \left(
X; \alpha \right)  \right\} \dfrac{
	2Z-1} {f(Z\mid X; \hat{\gamma}_{\text{\it mle}})} = 0,
\end{equation}
where $h$ and $g$ are vector functions of the right dimension, such as the identity function; $\hat{\gamma}_{\text{\it mle}}, \hat{\theta}_{\text{\it mle}}$ and $\hat{\iota}_{\text{\it mle}}$ are maximum likelihood estimators of $\gamma, \theta$ and $\iota$, respectively. 
It can be shown that $\hat{\beta}_{\text{\it dr}}$ is CAN in the union model of $\m_1$ and $\m_2$, and $\hat{\alpha}_{\text{\it dr}}$ is CAN in the union model of $\m_1$ and $\m_3$ \citep{robins2001comment}.
 
%However, we lose triple robustness as $E[EIF_{\Delta}]\neq 0$ if models for $f(Z\mid X)$ and $p_0^Y(X)$ are correct but models for $p_0^D(X), p_0^Y(X)$ and $\delta_D(X)$ are incorrect.
%The triple robustness of the efficient influence function, however, does not directly translate into triple robustness of estimators. In practice, estimation of \eqref{eqn:ee_unknown_general} typically relies on direct parametric models . 

% The proof of double robustness in the first case is given in  \cite{richardson2017modeling}. Similarly one can show double robustness for the second case. 
 
A multiply robust estimator $\hat{\Delta}_{mr}$ is given as follows:
 \begin{equation}
 \label{eqn:ee_known_general}
\hat{\Delta}_{mr} = \mathbb{P}_n \left\{  \dfrac{1}{\delta^D(X;\hat{\beta}_{\text{\it dr}})}\left(Y-D\delta \left( X; \hat{\alpha}_{\text{\it dr}} \right) -p_{0}^Y \left( X; \hat{\iota}_{\text{\it mle}} \right) +p_{0}^{D}(X; \hat{\theta}_{\text{\it mle}}) \delta \left(X; \hat{\alpha}_{\text{\it dr}} \right)   \right)  \dfrac{2Z-1 }{f(Z\mid X; \hat{\gamma}_{\text{\it mle}}) } +\delta \left( X;\hat{\alpha}_{\text{\it dr}} \right) \right\}.
 \end{equation}
Theorem \ref{thm:double_robustness} summarizes the key properties of $\hat{\Delta}_{mr}$. The proof is left to the on-line supplementary materials. 

\begin{theorem}
	\label{thm:double_robustness} 
	Under standard regularity conditions and the positivity assumption that both $\delta^D(X)$ and $f(Z\mid X)$ are bounded away from 0,
$\hat{\Delta}_{mr}$  is a CAN estimator of $\Delta$  in the union model of $\m_1$, $\m_2$, $\m_3$. Furthermore, if all  models $\m_1$, $\m_2$, $\m_3$ are correct, then the variance of $\hat{\Delta}_{mr}$ attains the semiparametric efficiency bound in the union model $\Mu$,  regardless of the choice of $h$ and $g$.
\end{theorem}

We now discuss the more challenging case of binary $Y$. As noted by \citet[][Remark 3.1]{richardson2017modeling},  multiple robustness is a useful property only if it is possible for models $\m_1$, $\m_2$, $\m_3$ to be correct a priori. Specifically, the model parameters in each of $\m_1$, $\m_2$, $\m_3$ need to be variation independent of each other. Hence when $Y$ is binary, rather than specifying models $p_0^D(X;\theta)$ and $p_0^Y(X;\iota),$ we assume models $OP^D(X;\eta)$  and $OP^Y(X;\zeta)$ instead.  As explained in Section \ref{sec:regression_estimator}, together with models $\delta^Y(X;\alpha)$ and $\delta^D(X;\beta)$, these odds product models 
%give rise to the likelihood $f(D\mid Z,X; \beta,\eta)$ and $f(Y\mid Z,X;\alpha,\beta,\zeta)$, and hence 
imply  $p_0^D(X;\beta, \eta)$ and $p_0^Y(X;\alpha, \beta, \zeta).$ 
Theorem \ref{thm:double_robustness} holds if in \eqref{eqn:ee_beta} -- \eqref{eqn:ee_known_general}, $p_0^D(X;\hat{\theta}_{\text{\it mle}})$ is replaced with $p_0^D(X;\hat{\beta}_{\text{\it 2mle}}, \hat{\eta}_{\text{\it 2mle}})$ and $p_0^Y(X;\hat{\iota}_{\text{\it mle}})$ is replaced with $p_0^Y(X;\hat{\alpha}_{\text{\it 2mle}}, \hat{\beta}_{\text{\it 2mle}}, \hat{\zeta}_{\text{\it 2mle}}).$

We also note that   $\hat{\Delta}_{mr}$ is (locally) efficient and multiply robust but may not be bounded. To remedy this, 
 we note that 
if we choose the first element of the vector function $g(X)$ to be $\dfrac{1}{\delta_D(X; \hat{\beta}_{\text{\it dr}})}$,
then 
%in the case of continuous $Y$, $\hat{\alpha}_{\text{\it dr}}$ solves
%$$
%		\mathbb{P}_n \dfrac{1}{\delta_D(X;\hat{\beta}_{\text{\it dr}})}\left\{ Y-D\delta \left( X; \alpha \right) -p_{0}^Y \left( X; \hat{\iota}_{\text{\it mle}} \right) +p_{0}^{D}(X; \hat{\theta}_{\text{\it mle}}) \delta \left(X; \alpha \right)  \right\} \dfrac{
%			2Z-1} {f(Z\mid X; \hat{\gamma}_{\text{\it mle}})} = 0,
%$$
%and  in the case of binary $Y$, 
$\hat{\alpha}_{\text{\it dr}}$ solves
$$
\mathbb{P}_n \dfrac{1}{\delta_D(X;\hat{\beta}_{\text{\it dr}})}\left\{ Y-D\delta \left( X; \alpha \right) -p_0^Y(X;\hat{\alpha}_{\text{\it 2mle}}, \hat{\beta}_{\text{\it 2mle}}, \hat{\zeta}_{\text{\it 2mle}})+p_0^D(X;\hat{\beta}_{\text{\it 2mle}})\delta \left(X; \alpha \right)  \right\} \dfrac{
	2Z-1} {f(Z\mid X; \hat{\gamma}_{\text{\it mle}})} = 0.
$$
Together with \eqref{eqn:ee_known_general}, this implies a bounded multiply robust estimator
\begin{equation*}
\label{eqn:ee_bounded}
\hat{\Delta}_{\text{\it b-mr}} = \mathbb{P}_n \delta(X;\hat{\alpha}_{\text{\it  dr}}).
\end{equation*}
%which is a plug-in estimator.  As before, when the parameter space for $\Delta$ is constrained, one may choose an appropriate model for $\delta(x)$ so that 
%the resulting estimator is bounded, (locally) efficient and multiply robust.  

The asymptotic variance formula of each estimator described in this section follows from standard M-estimation theory. Alternatively, bootstrapping methods may  be used for variance estimation in practice.

\begin{remark}
	In some contexts  interest may lie in estimating the conditional Wald estimand $\delta(x)$, especially when $X$ contains effect modifiers.  We note that given a model $\delta(X;\alpha),$ $\hat{\alpha}_{dr}$ that solves \eqref{eqn:ee_delta} is a doubly robust estimator, meaning that it is CAN  in the union of models $\m_1$ and $\m_3$.
\end{remark}

\begin{remark}
	Our construction of $\hat{\Delta}_{b-mr}$ is motivated by \cite{robins2007comment}. There are also other constructions of bounded estimators available for estimating the average treatment effect in the absence of unmeasured confounding; see for example, \cite{tan2010bounded} and \cite{vermeulen2015bias}.  In principle, these approaches  may also be applied in our context.
\end{remark}

\subsection{Discussions on multiple robustness}

Multiply robust estimators have been proposed previously in the literature.  
Prior to our work,  \cite{vansteelandt2007estimation} propose 2$^T$-multiply robust estimators in the context of longitudinal measurements with non-monotone missingness, 
 \cite{vansteelandt2008multiply} propose multiply robust estimators for statistical interactions, \cite{tchetgen2009commentary} proposes a multiply robust estimator to adjust for drop-out in randomized trials and 
  \cite{tchetgen2012semiparametric} propose multiply robust estimators for the marginal natural indirect and direct causal effects.   
  These estimators have in common that they are CAN 
  if the analyst correctly specifies the models for one, but not necessarily more than one components of the observed data law. As in our case,  these components may contain multiple elements with possible overlaps.

%
%All these multiply robust estimators build on the doubly robust estimators proposed by   \cite{robins1994estimation}. Specifically, consider estimating  the observed data functional $\Delta_g = E_X (E[Y\mid D=1,X] - E[Y\mid D=0,X]),$ which can be interpreted as $ATE$ under the ignorability assumption: $D\ind Y(d)\mid X, d=0,1$. \cite{robins1994correcting} proposed a doubly robust estimator $\hat{\Delta}_{DR}$ that remain consistent if either the outcome regression model $E[Y\mid D,X; \beta_{\text{\it or}}]$ or the propensity score model $P(D=1\mid X; \gamma_{\text{ps}})$ is correctly specified. 
%It is naturally to ask why we are able to obtain multiply robust estimators for our observed data functional $\Delta$. Intuitively, the number of robustnesses one can get depends on the observed data structure. In our problem, the observed data likelihood factorizes as
%\begin{equation}
%\label{eqn:fac_1}
%	f(Y,D,Z,X) = f(Y\mid Z,X)f(D\mid Z,X)  f(Z\mid X) f(X) c(Z,X),
%\end{equation}
%where $c(Z,X) = \dfrac{f(Y,D\mid Z,X)}{f(Y\mid Z,X)f(D\mid Z,X)}$ captures the association between $Y$ and $D$ conditional on $Z$ and $X$. In comparison, suppose one only observe $(D,X,Y)$ and wishes to estimate $\Delta_g = E_X (E[Y\mid D=1,X] - E[Y\mid D=0,X]),$ then the observed data likelihood factorizes as 
%\begin{equation}
%\label{eqn:fac_2}
%	f(Y,D,X) = f(Y\mid D,X) f(D\mid X) f(X).
%\end{equation}
%The right hand side of \eqref{eqn:fac_1} involves more functionals than the right hand side of \eqref{eqn:fac_2}, which partly explains  our triple robustness property.
	
We remark that the multiple  robustness result in Theorem \ref{thm:double_robustness} is non-trivial.   In particular, it relies on a novel parameterization of the efficient influence function in term of  functions $\delta(X), \delta^D(X), p_0^Y(X), p_0^D(X)$ and  $f(Z\mid X).$
The intuition for this parameterization comes  from Sections \ref{sec:regression_estimator} -- \ref{sec:g-estimation}, where we show that it is possible to construct CAN estimators of $\Delta$ in each of models $\m_1$, $\m_2$, $\m_3$. To see why our parameterization is important, consider an alternative parameterization of $EIF_\Delta$:
\begin{equation}
\label{eqn:eif_alter}
EIF_\Delta = \dfrac{2Z-1 }{f(Z\mid X) }\left( \dfrac{Y}{\delta^D
	\left( X\right) }-\dfrac{p_{0}^Y \left( X\right) }{%
	\delta^D \left( X\right) }-\dfrac{D}{(\delta^D \left( X\right))^2 }\delta^Y \left(
X\right)  +p_{0}^{D}(X) \dfrac{\delta^Y \left(
	X\right) }{(\delta^D \left( X\right))^2 } \right) +\dfrac{\delta^Y(X)}{\delta^D(X)} -\Delta,
\end{equation}
which is a function of $\delta^Y(X)$ rather than $\delta(X)$. The analogue of $\m_1$ is hence $\m_1^\prime$ that models for $\delta^Y(X), \delta^D(X), p_0^Y(X)$ and $p_0^D(X)$ are correct, and the analogue of $\m_3$ is hence $\m_3^\prime$ that  models for $f(Z\mid X)$ and $\delta^Y(X)$ are correct. 
Although $ EIF_\Delta$ has zero expectation when evaluated in the union of $\m_1^\prime$ and $\m_2$, the multiple robustness
property is less obvious with representation \eqref{eqn:eif_alter} as the expectation of $EIF_{\Delta}$ is not necessarily zero  in  model $\m_3^\prime$. More generally speaking,  given an observed data model and a parameter of interest, the efficient influence function is a
	unique random variable  \citep{bickel1998efficient}. However, different parameterizations of the same efficient influence function may lead to different conclusions about its robustness property. We refer interested readers to \cite{tchetgen2010doubly} for another example of this phenomenon. 
	%	Although these statements are all correct, some are more informative than the others. 
It remains an open problem  whether one can construct an estimator for $\Delta$ that are robust to models for more components of the likelihood.

We also clarify the conceptual difference between the multiple robustness property discussed in this article and the improved double robustness property discussed in some recent papers in the missing data literature \citep{han2013estimation,cefalu2017model,naik2016multiply}. These improved doubly robust estimators are constructed based on multiple working models for \emph{two} components of the likelihood: the outcome regression and the propensity score. They  remain CAN  if at least one of the working models is correct. 
As pointed out by \cite{julieta2017multiple} and \cite{li2017demystifying}, technically these estimators are still doubly robust rather than multiply robust, because they require specification of a large model for each of the \emph{two} components of the likelihood, each of them being the union of several smaller models.   In contrast, our multiply robust estimators allow for specification of models for \emph{multiple} components of the likelihood.

It is natural to ask whether one can construct estimators that enjoy the improved multiple robustness property in our context. In principle, the answer is positive. For example, suppose  $f_j(Z\mid X; \gamma_j),  j = 1\ldots, J$ are multiple parametric models for $f(Z\mid X)$ but only the first one is correct. To construct a consistent estimator for $f(Z\mid X)$ in this case, consider the parametric model 
$$
f(Z\mid X; \alpha_j, \gamma_j, j=1,\ldots, J) = \sum\limits_{j=1}^J \alpha_j f_j(Z\mid X; \gamma_j).
$$
One can estimate the model parameters in two steps. In the first step, one obtains the maximum likelihood estimates of $\gamma_j$, denoted as $\hat{\gamma}_j$. In the second step, one fits a linear regression of $Z$ on  $f_j(Z\mid X; \hat{\gamma}_j), j=1,\ldots, J$. 
It is not hard to see that $\hat{\alpha}_j \rightarrow_p 1(j=1)$ wherein $1(\cdot)$ is the indicator function, and ${f}(Z\mid X;  \hat{\alpha}_j, \hat{\gamma}_j, j=1,\ldots, J)$ estimates $f(Z\mid X)$ consistently. Estimators for other parts of the observed data law can be constructed in a similar way.

Finally, we emphasize that multiply robust estimation is possible as we do not commit to any observed data model in establishing identification. In contrast, the linear models \eqref{eqn:lsem1} and \eqref{eqn:lsem2}  imply models in  $\m_1$, so that the corresponding two-stage least square estimator may not be consistent outside of $\m_1$. Similarly, 	\cite{okui2012doubly} and \cite{vansteelandt2015robustness} also implicitly consider estimation of the 
%ATE under a sufficient condition for \cite{hernan2006instruments}'s identification assumption, that is $Y(d=1,U) - Y(d=0,U)$ does not depend on $U$. 
average Wald estimand, but they assume that a model for  $\delta(X)$ is  correctly specified. It follows that  their estimators are only  CAN in the union of $\m_1$ and $\m_3$ and therefore only doubly robust.

\section{Simulation studies}
\label{sec:simulation}

In this section, we evaluate the finite sample performance of the proposed estimators. In our simulations, the baseline covariates $X$ include an intercept  and a continuous variable $X_2$ uniformly distributed on the interval $(-1,-0.5) \cup (0.5, 1).$ The unmeasured confounder $U$ is Bernoulli distributed with mean 0.5.  Conditional on $X$ and $U$, the instrument $Z$, treatment $D$ and outcome $Y$ are generated from the following models:
\begin{flalign*}
	P(Z=1\mid X,U) &= \pi(X) = \text{expit}(\gamma^T X); \\
	\delta^D(X) &= \tanh(\beta^T X);\\
%	 \log \dfrac{P(D=1\mid Z=1, X)P(D=1\mid Z=0, X)}{P(D=0\mid Z=1, X)P(D=0\mid Z=0, X)} 
OP^D(X) &= \exp(\eta^T X); \\
	P(D=1\mid Z,X,U) &= p_0^D(X) + Z \delta^D(X) + \kappa (2U-1), \\
	\delta(X) &= \tanh(\alpha^T X);\\
		OP^Y(X) &= 
%		\log \dfrac{P(Y=1\mid Z=1, X)P(Y=1\mid Z=0, X)}{P(Y=0\mid Z=1, X)P(Y=0\mid Z=0, X)} 
\exp(\zeta^T X); \\
	P(Y=1\mid Z, X,U) &= p_0^Y(X) + Z \delta^D(X) \delta(X) + \kappa (2U-1),
\end{flalign*}
where  $p_0^D(X)$  is obtained from $\delta^D(X)$ and $OP^D(X)$ using  \eqref{eqn:p0d},  $p_0^Y(X)$ is obtained from $\delta^D(X), \delta(X)$ and $OP^Y(X)$ using  \eqref{eqn:p0y},  $\alpha=(0.1,0.5)^T, \beta=(0,-0.5)^T, \gamma=(0.1,-0.5)^T, \zeta = (0,-1)^T$, $\eta = (-0.5,1)^T$ and $\kappa = 0.1.$ 
We are interested in estimating the average Wald estimand $\Delta = E_X \delta(X),$ whose true value is 0.087.

We consider five estimators:
\begin{itemize}
	\item[{\sf {b-reg}}:] The bounded regression estimator placing models on $\delta(X),  \delta^D(X), OP^D(X)$ and $OP^Y(X).$
%	\item[{\sf {ipw}}:] The inverse probability estimator $\hat{\Delta}_{\text{\it ipw}}$
	\item[{\sf {b-ipw}}:] The bounded inverse probability estimator  solving equation \eqref{eqn:ee_ipw} with $h_1(X)=h_2(X) = X$.
		\item[{\sf {g}}:] The g estimator solving equation \eqref{eqn:g_est} with $h_3(X) = X.$
	\item[{\sf {mr}}:] The multiply robust estimator obtained by solving equations  \eqref{eqn:ee_beta},  \eqref{eqn:ee_delta} and \eqref{eqn:ee_known_general} with $h(X) = g(X) = X.$
	\item[{\sf {b-mr}}:] The bounded multiply robust estimator obtained by solving equations  \eqref{eqn:ee_beta},  \eqref{eqn:ee_delta} and \eqref{eqn:ee_known_general} with $h(X) = X, g(X) = \left(X_2, {1}/{\delta_D(X;\hat{\beta}_{dr})}\right).$ 
\end{itemize}

We also consider scenarios in which the models in $\m_1$, $\m_2$, $\m_3$ are misspecified. In these cases, instead of using $X$, the analyst uses covariates $X^\dagger$ that consists of an intercept and  $X_2^\dagger$; the latter covariate  is generated from an independent standard normal distribution. Specifically, in the following we report results from the following  five scenarios:
\begin{itemize} 
	\item[All correct:] $X$ is used in models $\m_1$, $\m_2$, $\m_3$;
	\item[$\m_1$ correct:]  $X$ is used in model $\m_1$, but $X^\dagger$ is used in the model for $P(Z=1\mid X)$;
	\item[$\m_2$ correct:] $X$ is used in model $\m_2$, but $X^\dagger$ is used in the models for $\delta(X), OP^D(X), OP^Y(X)$;
	\item[$\m_3$ correct:]  $X$ is used in model $\m_3$, but $X^\dagger$ is used in the models for $\delta^D(X), OP^D(X), OP^Y(X)$;
	\item[All wrong:] $X^\dagger$ is used in the models for $\delta(X), \delta^D(X), OP^D(X), OP^Y(X)$
	and $P(Z=1\mid X).$
\end{itemize}
Motivated by a reviewer's comment, in the on-line supplementary materials we report results for all 16 combinations of correct/incorrect
specifications of the following four sets of models: $\delta(X),\delta^D(X), f(Z | X)$ and $(OP^Y(X),OP^D(X))$.
In all these settings, the working model for computing {\sf {b-ipw}}  uses the same covariate as the model for $\delta^D(X)$. 

\begin{figure}
	% 3. Data and programming\cluster\submitted090516\1.6 Summary.R
	\centering
	\includegraphics[width=.75\textwidth]{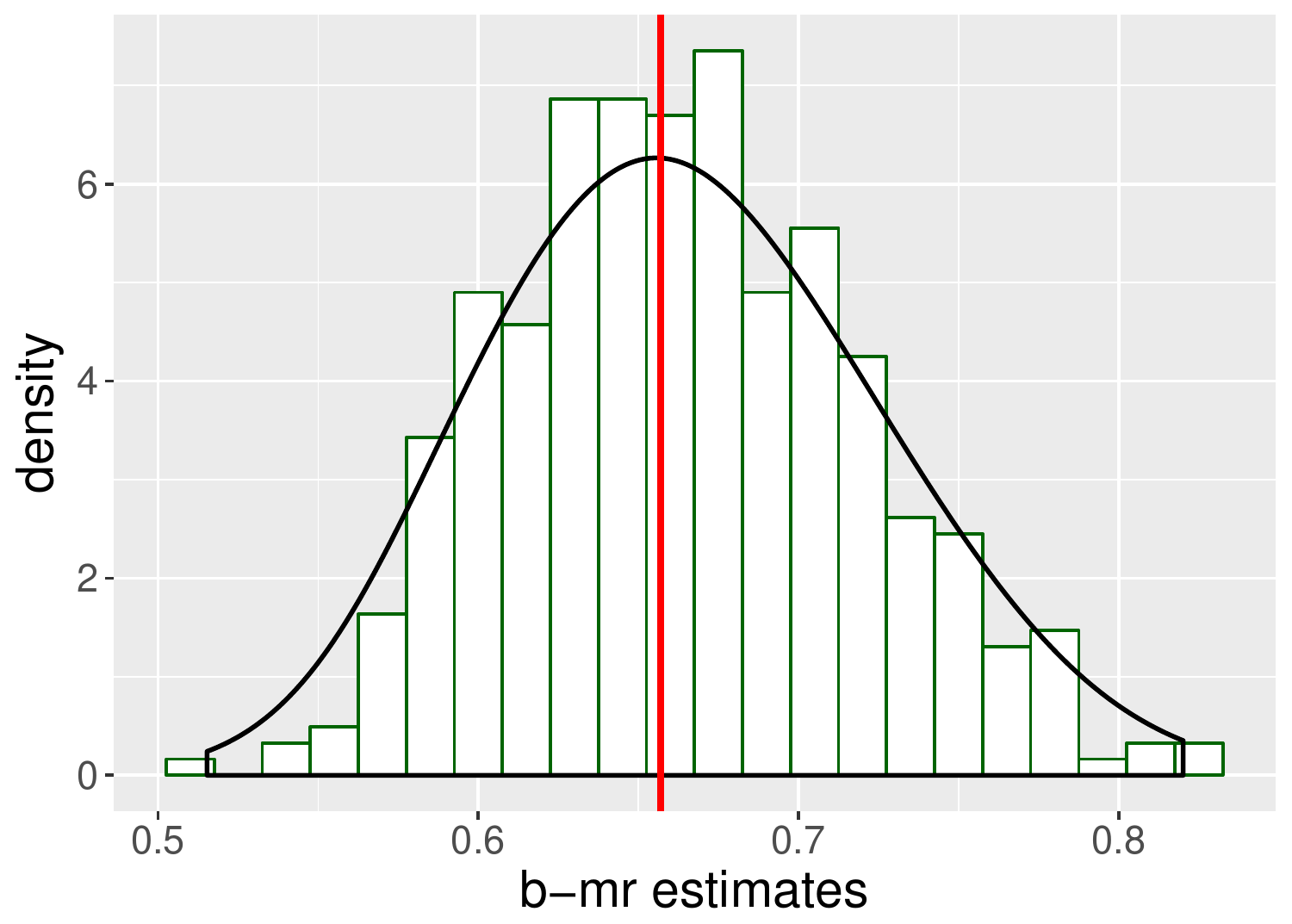}
	\caption{Estimates of {\sf b-mr} when {\sf mr} estimates go outside of [-1,1]. The red vertical line indicates the true value.}
	\label{fig:descriptive}
\end{figure}

All  simulation results are based on 1000 Monte-Carlo runs of $n=500$ units each.
Table \ref{tab:est} summarizes the simulation results. As predicted by our theoretical results, {\sf b-reg} has small bias only when model $\m_1$ is correct, {\sf b-ipw} has small bias only when model $\m_2$ is correct, {\sf g} has small bias only when model $\m_3$ is correct, while {\sf b-mr} has small bias when one, but not necessarily more than one of models $\m_1$, $\m_2$, $\m_3$ is correct. 
%Although  {\sf b-mr} is slightly biased upwards when all  models are correct, nonetheless the bias decreases with sample size. For example, the bias is smaller than 0.001 if the sample size increases to 5000; see Table \ref{tab:est2} in the on-line supplementary materials for details. 
%The variance of {\sf b-mr} is slightly larger than that of {\sf b-reg} and {\sf g}, suggesting that we pay a small price for triple robustness. 
We also note that {\sf mr} is more variable than  {\sf b-mr}, especially when model $\m_1$ and $\m_2$ are both misspecified. This is because under these scenarios, $\delta^D(X)$ is misspecified. Although in our simulation settings, the true value of $\delta^D(X)$ is bounded away from 0, under model misspecification the estimated value for $\delta^D(X)$ may be  close to 0, leading to instability in the naive multiply robust estimator. There is also no guarantee that {\sf mr} falls within the interval [-1,1]. In fact, when only $\m_3$ is correct, in which case {\sf mr} is supposed to be consistent, it produces an estimate outside of [-1,1] for 77.6\%  of the simulation samples. In contrast, Figure  \ref{fig:descriptive} plots the distribution of {\sf b-mr} for these 776 simulation samples. One can see that even for these challenging samples, {\sf b-mr} performs quite well. 

\begin{table}
	% 2.3 Summary.R - submitted050317
	\begin{center}
		\small
		\caption{Monte Carlo results of the proposed estimators under different simulation settings. The true value for the average Wald estimand is 0.087. The sample size is 500}
		\bigskip
		\label{tab:est}
		\begin{tabular}{cccrrrrrrr}
			\toprule
	     &\multicolumn{3}{c}	{ Model} &&  \multicolumn{5}{c}	{Estimator}   \\[2pt]
	     	\cmidrule(r){2-4} \cmidrule(l){6-10} 
		&	$\m_1$ & $\m_2$ & $\m_3$ && {\sf b-reg} & {\sf b-ipw}  & {\sf g} & {\sf mr}  & {\sf b-mr} \\
			\midrule
			\multicolumn{4}{l}{Bias(SE) \quad \quad\quad \quad}	& &&& & & \\[2pt]
&$\checkmark$  & $\checkmark$  & $\checkmark$   && 0.004(0.005) & 0.006(0.005) & 0.002(0.005) & 0.006(0.005) & 0.010(0.005) \\ 
&$\checkmark$  & $\times$  & $\times$  && 0.004(0.005) & 0.317(0.004) & 0.319(0.003) & 0.008(0.005) & $-$0.011(0.005) \\ 
&$\times$  & $\checkmark$  & $\times$  && 0.054(0.004) & 0.006(0.005) & 0.097(0.022) & 0.001(0.005) & 0.006(0.006) \\
&$\times$  & $\times$  & $\checkmark$  && 0.258(0.020) & $-$0.088(0.026) & 0.002(0.005) & 8.336(8.776) & 0.007(0.005) \\ 
&$\times$  & $\times$  & $\times$  && 0.294(0.020) & 0.088(0.024) & 0.290(0.021) & $-$98.261(98.916) & 0.162(0.020) \\  [10pt]
			\multicolumn{4}{l}{RMSE}	&& & & && \\[2pt]
&$\checkmark$  & $\checkmark$  & $\checkmark$ &&  0.143 & 0.157 & 0.146 & 0.151 & 0.153 \\
&$\checkmark$  & $\times$  & $\times$ &&  0.143 & 0.157 & 0.146 & 0.151 & 0.153 \\
&$\times$  & $\checkmark$  & $\times$ && 0.136 & 0.157 & 0.691 & 0.172 & 0.201 \\ 
&$\times$  & $\times$  & $\checkmark$ && 0.645 & 0.830 & 0.146 & 277.389 & 0.151 \\ 
&$\times$  & $\times$  & $\times$ && 0.617 & 0.759 & 0.659 & 3126.442 & 0.643 \\ 
			\bottomrule 
		\end{tabular}
	\end{center}
\end{table}

\section{The causal effect of education on earnings}
\label{sec:real_data}

%Use Abadie's 401(k) data set with binary outcome.
%
%(Previously I failed to reject his IV model.)

To illustrate the proposed methods, we reanalyze data from the National Longitudinal Survey of Young Men (NLS) \citep{card1995using,tan2006regression,okui2012doubly,wang2017falsification}, which consist of observations on 5525 men aged between 14 and 24 in 1966.  Among them, 3010  provided valid education and wage responses in the 1976 follow-up. We are interested in estimating the causal effect of education on earnings, which might be confounded by  unobserved preferences for education levels. \cite{card1995using} proposes to use presence of a nearby 4-year college as an instrument. Following \cite{tan2006regression}, we consider education beyond high school as the treatment. In this data set, 2053 (68.2\%) lived close to a 4-year college, and  1521 (50.5\%)  had education beyond high school.   To illustrate the proposed methods with binary outcomes, we follow \cite{wang2017falsification} to  dichotomize the outcome wage at its median, that is 5.375 dollars per hour. 

We adjust for age, race, father and mother's education levels, indicators for  residence  in the south and  a metropolitan area  and IQ scores, all measured in 1966.  
Among them, race, parents' education levels and residence are included as they may affect both the instrument and outcome; age is included as it is likely to modify the effect of education on earnings, and IQ scores, as a measure of underlying ability, are included as they may modify both the effect of proximity to college on education, and the effect of education on earnings. Following \cite{card1995using}, we use  mean imputation for missing values, and include indicators of imputation in the models.  The NLS was also not a representative sample of the US population. We account for this by reweighing observations using their sampling weights.  

We first apply the test of \cite{wang2017falsification} to check the proposed instrumental variable model. Analysis results show that the proposed IV model cannot be falsified with the observed data. We then apply the proposed methods to estimate the causal effect of education on earnings. In addition to the estimators included in Section  \ref{sec:simulation}, we report the crude estimate which does not account for any confounding, and
the Risk Difference Regression (RDReg) estimate \citep{richardson2017modeling}, which only accounts for observed confounders; the RDReg estimate was obtained using {\tt R} package {\tt brm} \citep{wangbrm}. We also report the two-stage least square (2SLS) estimate, which is believed to approximate the causal relation of interest \citep{angrist2008mostly} despite not respecting the natural model constraints  of a binary outcome.   The results are summarized in Table \ref{tab:result}, wherein the confidence intervals are obtained by quantile-based nonparametric bootstrap. 

	\begin{table} 
		% submitted110616 2.2. Summary.R, submitted110816 2.2 Summary.R
		\begin{center}
			\caption{Estimates of the effect of education beyond high school on  earnings (dichotomized at median) } 
			\bigskip
			\begin{tabular}{rrrrrrrrrr}
				\toprule
				Method & Point Estimate  &  95\% Confidence Interval \\
				\midrule
				{\sf naive} & 0.122 & (0.085,0.162) \\[2pt]
				{\sf RDReg} & 0.037 &  ($-$0.009,0.080)                \\ [2pt]
				{\sf 2SLS} &  0.469 & ($$-$$0.140,1.327) \\[2pt] 
				{\sf b-reg} & 0.849 & (0.239,0.978) \\ 
				{\sf b-ipw} & 0.424 & ($-$1.000,1.000) \\ 
				{\sf g} & 0.079 & ($-$0.355,1.000) \\ [2pt]
				{\sf mr} & 0.328 & ($-$60.394,65.339) \\ 
				{\sf b-mr} &  0.344 & ($-$0.373,0.938) \\ 
				\bottomrule
			\end{tabular}
			\label{tab:result}
		\end{center}
	\end{table}

Table \ref{tab:result} summarizes the results. The bootstrapped estimates are based on 1000 bootstrap samples. Both  2SLS and bounded multiply robust estimation yield  substantially higher point estimates than  Risk Difference Regression, suggesting that unmeasured confounding leads to a downward bias in the regression estimate of causal effect; this is consistent with previous findings using the log wage as the outcome \citep{card1995using,okui2012doubly}. 
Unlike 2SLS, the confidence interval given by the bounded multiply robust estimation is contained within the interval [$-$1, 1], confirming boundedness of the proposed estimator. 
The bounded IV regression estimate is very close to 1, which is unlikely given that the outcome takes value 0 or 1. This is probably due to misspecification of one or more  models contained in $\m_1$. Despite this possible misspecification, the multiply robust methods still yield reasonable estimates. This indicates that  multiply robust estimation does provide some protection against model misspecification.  Furthermore, the multiply robust estimates are close to the bounded IPW estimate. This suggests that the models in $\m_2$, that are $\delta^D(X;\beta)$ and $f(Z\mid X;\gamma)$, may be close to the truth \citep{tchetgen2010semiparametric}.

\section{Discussion}
\label{sec:discussion}
%
%Extensions to discrete IV: the connection between the Wald estimator and the grouped linear regression \citep[][\S 4.1.3]{angrist2008mostly}
%
%Also in principle our approach works for continuous Z as well just choose two levels z and z* (say z+1) instead of 0 and 1  then everything goes through but then you would probably want to integrate over z and z*, very strange because causal effect itself is independent of Z.

IV methods are widely used to identify causal effects in the presence of unmeasured confounding. In practice, 
applied researchers are primarily interested in estimating the ATE.  In contrast, the majority of statistical IV literature  focuses on estimating the LATE, a legitimate causal parameter but often of secondary interest. We address this discrepancy by proposing novel  assumptions that allow for identification of the ATE. We have argued that our novel identification assumptions are often more plausible than previous assumptions as they do not place constraints on the observed data distribution and do not suffer from the same limitations as previous identification approaches. Nevertheless, our identification assumptions lead to the same observed data functional as previous methods targeting the ATE, so that our proposed estimators can also be used to estimate the ATE under previous identification frameworks. 

In this paper we discuss in detail how to obtain bounded estimators  of the additive ATE with binary outcomes. This is especially relevant as most previous semiparametric IV methods primarily deal with continuous outcomes.  
Correspondingly, it is of interest to develop bounded estimators of the LATE with binary outcomes. Our approach in this paper can also be extended to improve causal effect estimation  in the case of the multiplicative ATE, as well as to the context of a failure time outcome under additive or multiplicative hazards models.   Finally, we have focused on the case of binary instrument and treatment in this paper. Extension of the proposed methodology to the case of general instrument or treatment is an interesting topic for future research. 

%{\bf Do we need to include extensions to discrete/continuous instruments here?}

%generalize our approach to time varying setting, today's developments seem to suggest that we may be able to do so, this would potentially lead to estimation of an MSM with an IV

\section*{Acknowledgments}
The research was supported by  U.S. National Institutes of Health grants AI113251, ES020337 and AI104459. We thank the Associate Editor, two referees, Andrea Rotnitzky and Xu Shi for helpful comments.

\section*{Appendix}
\renewcommand{\thesubsection}{\Alph{subsection}}

\subsection{Proof of Theorem \ref{thm:identification}}
\label{appendix:wald_interpretation}

To simplify notation, conditioning on $X$ is implicit in our proof. We first note that  
	\begin{small}
		\begin{flalign*}
		\hspace*{-0.4cm}\delta^Y &= 	 \sum\limits_{z=0,1}  (2z-1) E[Y\mid Z=z]  \\
		&= \sum\limits_{z=0,1}  (2z-1) E_U E[Y\mid Z=z, U] \quad (Z \ind U) \\
       &= \sum\limits_{z=0,1} (2z-1) E_U \left\{ E[Y D\mid Z=z,U] + E[Y(1-D)\mid Z=z,U] \right\} \\
		&= E_U \sum\limits_{z=0,1} (2z-1) \left\{ E[Y(1)D \mid Z=z, U] +  E[Y(0)(1-D)\mid Z=z, U] \right\}  \\
   	&=E_U  \sum\limits_{z=0,1} (2z-1) \{  E[Y(1)\mid  Z=z,U]P(D=1\mid Z=z,U) + E[Y(0)\mid Z=z,U] P(D=0\mid Z=z,U)  \} 
		\ \ \ (Y(d) \ind D \mid (Z,U)) \\
			&=E_U  \sum\limits_{z=0,1} (2z-1) \{  E[Y(1)\mid  U]P(D=1\mid Z=z,U) + E[Y(0)\mid U] P(D=0\mid Z=z,U)  \} 
			\ \ \ (Y(d) \ind Z \mid U) \\
			&= E_U \left( E[Y(1)-Y(0)\mid U] \left\{E[D\mid Z=1, U] - E[D\mid Z=0, U]  \right\} \right).
			\numberthis\label{eqn:key}
%		&= E_U \sum\limits_{z=0,1} (2z-1) \{ E[Y(1)\mid  U]P(D(z)=1\mid U) + E[Y(0)\mid U] - E[Y(0)\mid U] P(D(z)=1\mid U)  \} \\
%		&= E_U \sum\limits_{z=0,1} (2z-1) P(D(z)=1\mid U)  \{  E[Y(1)\mid  U] - E[Y(0)\mid U]  \} \\
%		&=E_U \left(P(D(1)=1\mid U) - P(D(0)=1\mid U)\right)  E[Y(1)-Y(0)\mid  U].
		\end{flalign*}
	\end{small}
Under A5.a,
	\begin{equation*}
%	\label{eqn:a4a}
		E[D\mid Z=1,U] - E[D\mid Z=0,U]  = E[D\mid Z=1] - E[D\mid Z=0] = \delta^D;
	\end{equation*}
		under A5.b,
	\begin{equation*}
%	\label{eqn:a4b}		
	E[Y(1) - Y(0)\mid U]  = E[Y(1) - Y(0)].
	\end{equation*}
	and due to $Z\ind U$, $E_U \left\{E[D\mid Z=1, U] - E[D\mid Z=0, U]  \right\} = E[D\mid Z=1]-E[D\mid Z=0] = \delta^D.$
	Hence under either A5.a or A5.b, $\delta^Y = \delta^D E[Y(1)-Y(0)].$ This concludes our proof.

	\subsection{Interpretation of the average Wald estimand when A5.a and A5.b both fail}
	\label{appendix:wald}
	
	\begin{proposition}
		\label{prop:wald}
		Suppose $U$ is univariate and that for all $X$, $E[Y(1)-Y(0)\mid U,X]$ and \\ $\left(E[D\mid Z=1,U,X]-E[D\mid Z=0,U, X]\right) / \left(E[D\mid Z=1,X]-E[D\mid Z=0,X] \right)$ are both  non-decreasing/ non-increasing in $U$. Then the average Wald estimand
		$$
					\Delta \geq ATE.
		$$
		Furthermore, if $U$ is multivariate, then the conclusion still holds if the components of $U$ are conditionally independent given $X$.
	\end{proposition}
	
	The proof follows by noting that \eqref{eqn:key} implies 
	\begin{flalign*}
			\delta(X) - ATE(X) &= \dfrac{\delta^Y(X)}{\delta^D(X)} - ATE(X) \\
			&= \dfrac{1}{\delta^D(X)} \left( E_{U\mid X} \left( E[Y(1)-Y(0)\mid U,X] \left\{E[D\mid Z=1, U,X] - E[D\mid Z=0, U,X]  \right\} \right) - \right. \\
			&\left. E_{U\mid X} \left( E[Y(1)-Y(0)\mid U,X] \right) E_{U\mid X}  \left\{E[D\mid Z=1, U,X] - E[D\mid Z=0, U,X]  \right\}   \right) \\
			&= Cov_{U\mid X}\left( E[Y(1)-Y(0)\mid U,X], \dfrac{1}{\delta^D(X)}\left\{E[D\mid Z=1, U,X] - E[D\mid Z=0, U,X]\right\}\right).
	\end{flalign*}

	\thispagestyle{empty}
	\bibliographystyle{apalike}
	\bibliography{causal}

	\thispagestyle{empty}
	\bibliographystyle{apalike}
	\bibliography{causal}

	\clearpage
	
	\begin{center}
		%	\textit{Original Article}\bigskip
		
		{\LARGE Supplementary Materials for}
		
		{\LARGE ``Bounded, efficient and multiply robust  estimation of}
		
		{\LARGE  average treatment effects using instrumental variables''}
		
		{\Large Linbo Wang and Eric J. Tchetgen Tchetgen } $\ $
		
		{\large Department of Biostatistics}
		
		{\large Harvard University}
	\end{center}
	%%%%%%%%%% Merge with supplemental materials %%%%%%%%%%
	%%%%%%%%%% Prefix a "S" to all equations, figures, tables and reset the counter %%%%%%%%%%
	\setcounter{equation}{0}
	\setcounter{figure}{0}
	\setcounter{table}{0}
	\setcounter{page}{1}
	\makeatletter
	\renewcommand{\theequation}{S\arabic{equation}}
	\renewcommand{\thefigure}{S\arabic{figure}}
	\renewcommand{\thetable}{S\arabic{table}}
	\setcounter{section}{0}
	%%%%%%%%%% Prefix a "S" to all equations, figures, tables and rese t the counter %%%%%%%%%%

		\begin{abstract}
		This on-line supplementary file contains proofs of propositions and theorems in the main paper, as well as additional simulation results.
	\end{abstract}

	\section{Proof of Proposition 1}
	To simplify notation, conditioning on $X$ is  implicit in our proof.
	It was shown in \cite{bonet2001instrumentality} that (3) characterizes all the constraints on the observed data law implied by the canonical IV assumptions A1$^\prime$ -- A4. It is trivial to see that (3) do not impose restrictions on $f(Z).$ To show that they do not impose restrictions on $(P(Y=1\mid Z=z), P(D=1\mid Z=z)) \equiv (p^Y_z, p_z^D)$, we need to show for any $p^Y_{z}, p^D_{ z} \in [0,1],$ there exists  $(P(Y=y, D=d\mid Z=z), z,d,y=0,1)$ that satisfies (3). Let $x_z = P(Y=1,D=1\mid Z=z),$ then the constraints (3) can be rewritten as the following inequalities:
	\begin{flalign*}
	x_1 + p_0^D - x_0 &\leq 1, \\
	p_1^Y - x_1 +  1 + x_0 - p_0^Y - p_0^D & \leq 1,\\
	p_1^D - x_1 + x_0 &\leq 1, \\
	1+x_1 - p_1^Y - p_1^D + p_0^Y - x_0 &\leq 1,
	\end{flalign*}
	which are equivalent to the following constraint on $x_1-x_0$:
	\begin{equation}
	\label{eqn:cons_x}
	\max(p_1^Y - p_0^Y - p_0^D, p_1^D-1) \leq x_1-x_0 \leq		\min(1-p_0^D, p_{1}^Y + p_1^D - p_0^Y).
	\end{equation}
	It is not hard to verify that $\max(p_1^Y - p_0^Y - p_0^D, p_1^D-1)\leq		\min(1-p_0^D, p_{1}^Y + p_1^D - p_0^Y)
	$ so there always exist $(x_1, x_0)$ that satisfy \eqref{eqn:cons_x}. This completes the proof for claim (i). To show (ii), note that A5.a or A5.b holds trivially if 
	\begin{equation}
	\label{eqn:u_ind}
	U \ind (Z,D,Y(1), Y(0),X)
	\end{equation}
	and that \eqref{eqn:u_ind} is not falsifiable from the observed data.

	\section{Proof of Proposition 2}
	
	The proof follows from noting the following results:
	\begin{enumerate}
		\item $\delta^Y(X) = \delta^D(X) \delta(X) \in (-1,1)$;
		\item The mapping $(\delta^D(X), OP^D(X))$ is a diffeomorphism from $(-1,1)\times \mathbb{R}^+$ to $(0,1)^2$;
		\item The mapping $(\delta^Y(X), OP^Y(X))$ is a diffeomorphism from $(-1,1)\times \mathbb{R}^+$ to $(0,1)^2$;
		\item $P(Y=1\mid Z,X)$ is variation independent of $P(D=1\mid Z,X).$
	\end{enumerate}
	The second and third steps follow from results in \cite{richardson2017modeling}.

	\section{Proof of Theorem 2}
	
	Using standard theory on likelihood-based inference, one can show that under $\m_1$, $\hat{\alpha}_{\text{\it 2mle}}$ is asymptotically linear. 
	With slight abuse of notation, we also use $\alpha$ to denote  the true value of the parameter $\alpha$.
	Suppose that 
	$$
	\hat{\alpha}_{\text{\it 2mle}} = \alpha + \mathbb{P}_n IC_{\alpha} + o_p\left(1/{\sqrt{n}}\right),
	$$
	where $IC_\alpha$ is the influence curve of $\hat{\alpha}_{\text{\it 2mle}}.$ 
	We then have 
	\begin{flalign*}
	\mathbb{P}_n 	\delta(X;\hat{\alpha}_{\text{\it 2mle}}) &= 	\mathbb{P}_n\left\{ \delta(X;\alpha) + \dfrac{\partial \delta(X; \alpha)}{\partial \alpha} (\hat{\alpha}_\text{\it 2mle} - \alpha) \right\} + 	\mathbb{P}_n  o_p(\hat{\alpha}_\text{\it 2mle} - \alpha) \\
	&= \Delta + \mathbb{P}_n IC_\Delta  +\mathbb{P}_n \left\{ \dfrac{\partial \delta(X; \alpha)}{\partial \alpha} \left(   \mathbb{P}_n IC_{\alpha} + o_p(1/{\sqrt{n}})   \right) \right\} + o_p(1/\sqrt{n})  \\
	&=\Delta+ \mathbb{P}_n IC_\Delta  +\left\{ \mathbb{P}_n  \dfrac{\partial \delta(X; \alpha)}{\partial \alpha} \right\}  \left(   \mathbb{P}_n IC_{\alpha}\right)    + o_p(1/\sqrt{n})  \\
	&=\Delta+ \mathbb{P}_n IC_\Delta  + E \left\{ \dfrac{\partial \delta(X; \alpha)}{\partial \alpha} \right\}  \left(   \mathbb{P}_n IC_{\alpha}\right)  + O_p(1//\sqrt{n}) O_p(1//\sqrt{n})   + o_p(1/\sqrt{n})  \\
	&= \Delta+ \mathbb{P}_n \left\{ IC_\Delta   + E \left\{ \dfrac{\partial \delta(X; \alpha)} {\partial \alpha} \right\}  IC_\alpha \right\}  +  o_p(1/\sqrt{n}),
	\end{flalign*}
	where $IC_\Delta = \delta(X;\alpha) - \Delta$, and in the second and third steps above, we use the fact that for a given $n$, $\hat{\alpha}_{\text{\it 2mle}}$ does not depend on the specific unit $i$.

	\section{Proof of Theorem 3}
	
	We note that regardless of whether  or not  $\delta(X;\alpha)$ is correct, under $\m_2$ we have that both ($\hat{\beta}_{\text{\it ipw}},\hat{\Delta}_{\text{\it ipw}}$) and $(\hat{\beta}_{\text{\it ipw}},\hat{\Delta}_{\text{\it b-ipw}})$ solve equation (8) and the following  equation:
	\begin{equation}
	\label{eqn:Delta}
	\mathbb{P}_n \dfrac{Y}{\delta^D(X;\beta)} \dfrac{2Z-1}{f(Z\mid X;\hat{\gamma}_{\text{\it mle}})} - \Delta = 0,
	\end{equation}
	wherein the maximum likelihood estimator $\hat{\gamma}_{\text{\it mle}}$ is consistent for $\gamma.$
	
	Replacing $\hat{\gamma}_{mle}$ with $\gamma$,   equation (8) is an unbiased estimating equation as
	\begin{flalign*}
	&\quad E \left[h_1(X) \left( \dfrac{D(2Z-1)}{f(Z\mid X;\gamma)} - \delta^D(X;\beta)  \right)  \right] \\
	&=E_X h_1(X)  E \left[\left. \dfrac{D(2Z-1)}{f(Z\mid X;\gamma)} - \delta^D(X;\beta) \right| X  \right]  \\
	&= E_X h_1(X) (\delta^D(X) - \delta^D(X;\beta)) = 0,
	\end{flalign*}
	where the last line is due to the following identify:
	\begin{equation}
	\label{identity_3} \forall f, f(Z=1,X) - f(Z=0,X) = E\left[ \left. f(Z,X) \dfrac{2Z-1}{f(Z\mid X)} \right| X  \right].
	\end{equation}
	Similarly, replacing $\hat{\gamma}_{mle}$ with $\gamma$,   equation \eqref{eqn:Delta} is an unbiased estimating equation as because
	\begin{flalign*}
	&\quad E \left[ \dfrac{Y}{\delta^D(X;\beta)} \dfrac{2Z-1}{f(Z\mid X;\gamma)} \right] - \Delta \\
	&=E_X  E \left[\left.  \dfrac{Y}{\delta^D(X;\beta)} \dfrac{2Z-1}{f(Z\mid X;\gamma)} \right| X  \right] -\Delta \\
	&= E_X  \dfrac{\delta^Y(X)}{\delta^D(X)} - \Delta = 0.
	\end{flalign*}
	We now show that the equation (8) has a unique solution in the limit provided that i) the distribution of $X$ is non-degenerate; ii) $h_1(X)=X$. To see this, note that under these conditions,
	$$
	\dfrac{\partial}{\partial \beta}  E \left[h_1(X) \left( \dfrac{D(2Z-1)}{f(Z\mid X;\gamma)} - \delta^D(X;\beta)  \right)  \right] = \dfrac{4}{(e^{2\beta^T X}+1)^2} X^T X
	$$
	is positive definite. One can also see from this proof that any choice of $h_1(X)$ that makes $(h_1(X))^T X$ positive definite would suffice.
	
	The rest of the proof follows from standard M-estimation theory.

	\section{Proof of Theorem 5}

	In the following proof, we will use \eqref{identity_3} and  the following identities repeatedly:
	\begin{flalign}
	\label{identity_1}	& \forall f, E[f(X)S(Y\mid X)]  = 0; \\
	\label{identity_2}	& \forall f, E[f(X)(Y-E[Y\mid X])]  = 0,
	\end{flalign}
	where $S(Y\mid X)$ is the score function corresponding to the  conditional law of  $Y$ given $X$.

	To find the efficient influence function (EIF) for $\Delta$ in the union model $\Mu$, we first need to  find a gradient in $\Mu$. To do so, we  find the canonical gradient $G$ for $\Delta$ in the nonparametric model
	$\m_{\text{\it non}}$.  Specifically, we aim to find a random variable $G$, such that $E[G]=0$ and for all one-dimensional parametric submodels of $\m_{\text{\it non}}$, denoted as $f(Y,D,Z,X;t) = f(Y,D\mid Z,X;t) f(Z\mid X;t) f(X;t)$, we have 
	\begin{equation}
	\label{eqn:ic}
	\left.	\dfrac{\partial}{\partial t} \Delta_t \right|_{t=0} = E[G \cdot S(Y,D,Z,X;t)]|_{t=0},
	\end{equation}
	where $S(Y,D,Z,X;t) = \partial \log f(Y,D,Z,X;t) / \partial t$ and 
	$$
	\dfrac{\partial}{\partial t} \Delta_t  = \dfrac{\partial}{\partial t}  \int  \dfrac{ \int y \ dF_t(y\mid Z=1, X) - \int y \ dF_t(y\mid Z=0, X)}{\int d \ dF_t(d\mid Z=1, X) - \int d \ dF_t(d\mid Z=0, X)} f_t(X) \ dX   
	= \dfrac{\partial}{\partial t} E_{t}\left\{ \dfrac{\delta^Y_{t}\left(X\right) }{\delta^D_t\left( X\right) }\right\}.
	$$

	First note that
	\begin{eqnarray*}
		\left.	\dfrac{\partial}{\partial t}\Delta _{t} \right|_{t=0} &=& \left. \dfrac{\partial}{\partial t}E_{t}\left\{ \dfrac{\delta^Y_{t}\left(
			X\right) }{\delta^D_t\left( X\right) }\right\}  \right|_{t=0}\\
		&=& \left. E\left\{ \dfrac{\delta^Y \left( X\right) }{\delta^D \left( X\right) }%
		S(X)\right\} + E\left\{ \dfrac{\dfrac{\partial}{\partial t}\delta^Y _{t}\left( X\right) \delta^D \left(
			X\right) -\delta^Y \left( X\right) \dfrac{\partial}{\partial t}\delta^D _{t}\left( X\right) }{%
			(\delta^D\left( X\right))^2 }\right\}  \right|_{t=0}\\
		&=&\left. E\left\{ \dfrac{\delta^Y \left( X\right) }{\delta^D \left( X\right) }%
		S(Y,D,Z,X)\right\} + E\left\{ \dfrac{\dfrac{\partial}{\partial t}\delta^Y _{t}\left( X\right) \delta^D \left(
			X\right) -\delta^Y \left( X\right) \dfrac{\partial}{\partial t}\delta^D _{t}\left( X\right) }{%
			(\delta^D\left( X\right))^2}\right\}\right|_{t=0} \text{\quad(following \eqref{identity_1})},   
	\end{eqnarray*}%
	where $S(X)$ denotes the score function for $X$, and
	\begin{flalign*}
	\dfrac{\partial}{\partial t} E_t(Y\mid Z=z, X) &= \dfrac{\partial}{\partial t} \int y f_t(y\mid Z=z,X) dy \\
	&={ E\left[Y \dfrac{\dfrac{\partial}{\partial t} f_t(Y\mid Z=z, X)}{f_t(Y\mid Z=z, X)}   \right]    } \\
	&= E\left[(Y-E[Y\mid Z=z,X]) S(Y\mid Z=z,X)  \right]    \text{\quad(following \eqref{identity_1})}\\
	&= E\left[(Y-E[Y\mid Z=z,X]) S(Y, Z=z\mid X)  \right]    \text{\quad(following \eqref{identity_2})}.
	%&= \dfrac{\partial}{\partial t} \dfrac{ \int y f_t(y, Z=1\mid X) dy } { f_t(Z=1\mid X)  }  \\
	%&= \dfrac{ \int y f_t^\prime(y,Z=1\mid X) dy \times  f_t(Z=1\mid X) -  \int y f_t(y,Z=1\mid X) dy  \times f_t^\prime(Z=1\mid X)   }{ f_t^2(Z=1\mid X) } \\
	%&= \int y S(y, Z=1\mid X) f(y\mid Z=1, X) dy -  \int y f_t(y\mid Z=1,X) dy \times  S(Z=1\mid X) \\
	%&= E[Y S(Y, Z=1\mid X) \mid Z=1,X  ] - E(Y\mid Z=1, X) S(Z=1\mid X)  \\
	%&= E[Y S(Y,D, Z=1\mid X) \mid Z=1,X  ] - E(Y\mid Z=1, X) S(Y,D, Z=1\mid X)  \\
	%&= E[(Y-E(Y\mid Z=1,X)) S(Y,D,Z=1\mid X) ].
	\end{flalign*}
	%	\begin{flalign*}
	%	\dfrac{\partial}{\partial t} E(Y\mid Z=0, X) &= \dfrac{\partial}{\partial t} \int y f_t(y\mid Z=0,X) dy \\
	%	&= \dfrac{\partial}{\partial t} \dfrac{ \int y f_t(y, Z=0\mid X) dy } { f_t(Z=0\mid X)  }  \\
	%%	&= \dfrac{ \int y f_t^\prime(y,Z=0\mid X) dy \times  f_t(Z=0\mid X) -  \int y f_t(y,Z=0\mid X) dy  \times f_t^\prime(Z=0\mid X)   }{ f_t^2(Z=0\mid X) } \\
	%%	&= \int y S(y, Z=1\mid X) f(y\mid Z=1, X) dy -  \int y f_t(y\mid Z=1,X) dy \times  S(Z=1\mid X) \\
	%	&= E[Y S(Y, Z=0\mid X) \mid Z=0,X  ] - E(Y\mid Z=0, X) S(Z=0\mid X)  
	%	\end{flalign*}	
	%	
	Hence	
	\begin{flalign*}
	\dfrac{\partial}{\partial t}\delta^Y _{t}\left( X\right)  &= \dfrac{\partial}{\partial t} \left\{ E_t(Y\mid Z=1,X) - E_t(Y\mid Z=0,X) \right\} \\
	&=  E\left[(Y-E[Y\mid Z=1,X]) S(Y, Z=1\mid X)  \right] -  E\left[(Y-E[Y\mid Z=0,X]) S(Y, Z=0\mid X)  \right] \\
	%			&= f(Z=1\mid X) \dfrac{1}{f(Z=1\mid X)} E\left[(Y-E[Y\mid Z=1,X]) S(Y, Z=1\mid X)  \right] - \\
	%		&	f(Z=0\mid X) \dfrac{1}{f(Z=0\mid X)} E\left[(Y-E[Y\mid Z=0,X]) S(Y, Z=0\mid X)  \right] \\
	&= E\left(\left. \dfrac{2Z-1}{%
		f(Z\mid X) } \left( Y-E\left( Y|Z,X\right) \right)
	S(Y,Z\mid X)  \, \right| \, X \right) \text{\quad(following \eqref{identity_3})}  \\
	&= E\left(\left. \dfrac{2Z-1}{%
		f(Z\mid X) } \left( Y-E\left( Y|Z,X\right) \right)
	S(Y,Z,X)  \, \right| \, X \right) \text{\quad(following \eqref{identity_2})} \\ 
	&= E\left(\left. \dfrac{2Z-1}{%
		f(Z\mid X) } \left( Y-E\left( Y|Z,X\right) \right)
	S(Y,D,Z,X)  \, \right| \, X \right) \text{\quad(following \eqref{identity_1})}.
	\end{flalign*}
	Similarly, 
	$$
	\dfrac{\partial}{\partial t}\delta^D _{t}\left( X\right)  =E\left( \left. \dfrac{2Z-1}{%
		f(Z\mid X) } \left( D-E\left( D|Z,X\right) \right)
	S(Y,D,Z,X)  \, \right| \, X \right). 
	$$
	
	It then follows that 
	\begin{eqnarray*}
		\tilde{G}	&=&\dfrac{2Z-1}{f(Z\mid X) }\left( \dfrac{Y-E\left(
			Y|Z,X\right) }{\delta^D \left( X\right) }-\dfrac{\left( D-E\left( D|Z,X\right)
			\right) }{\delta^D \left( X\right) }\dfrac{\delta^Y \left( X\right) }{\delta^D
			\left( X\right) }\right)+\dfrac{\delta^Y \left( X\right) }{\delta^D \left( X\right) }  \\
		&=&\dfrac{2Z-1 }{f(Z\mid X) }  \dfrac{1}{\delta^D(X)}    \left\{ {Y-E\left(
			Y|Z,X\right) }-{\left( D-E\left( D|Z,X\right)
			\right) }\delta \left( X\right)\right\} +\delta \left(
		X\right)   \\
		&=&\dfrac{2Z-1 }{f(Z\mid X) }  \dfrac{1}{\delta^D(X)}    \left\{
		{Y}-\delta \left( X\right) \delta^D(X) Z-{p_{0}^Y \left( X\right) }-{D}\delta \left(
		X\right) +Z\delta^D \left( X\right) \delta(X) +p_{0}^{D}(X) {\delta \left(
			X\right) } \right\} +\delta \left( X\right)  \\
		&=& \dfrac{2Z-1 }{f(Z\mid X) } \dfrac{1}{\delta^D(X)}    
		\left\{ {Y}-{p_{0}^Y \left( X\right) }-{D}\delta \left(
		X\right)  +p_{0}^{D}(X) {\delta \left(
			X\right) } \right\} +\delta \left( X\right)
	\end{eqnarray*}%
	satisfies \eqref{eqn:ic}, in which we use the identities  $E\left( Y|Z,X\right) =\delta \left( X\right) \delta^D \left(
	X\right) Z+ p_{0}^Y \left( X\right) $ and $E\left( D|Z,X\right) =\delta^D \left(
	X\right) Z+ p_{0}^{D} \left( X\right) .$ Following  \eqref{identity_3} one can show that $E[\tilde{G}] = \Delta$. Consequently,  the canonical gradient in $\m_{\text{\it non}}$ is $G = \tilde{G} - \Delta$, which equals the EIF evaluated at observed data $(Y,D,Z,X)$, as shown by standard semiparametric efficiency theory \citep{bickel1998efficient}.  Results in \cite{robins2001comment} then show that the EIF in the union model $\Mu$ coincides with the EIF in $\m_{\text{\it non}}$. This concludes our proof.

	\section{Proof of Theorem 6}
	
	Under suitable regularity conditions \citep{white1982maximum}, the estimators $\hat{\alpha}_{\text{\it dr}}, {\hat{\beta}_{\text{\it dr}}}, \hat{\gamma}_{\text{\it mle}}, \hat{\theta}_{\text{\it mle}}, \hat{\iota}_{\text{\it mle}}$ converge in probability to fixed constants ${\alpha}^{\ast}, {\beta}^{\ast}, {\gamma}^{\ast}, {\theta}^{\ast}, {\iota}^{\ast}$ regardless of whether the corresponding models are correct or not. In the following, we denote $\delta(X; {\alpha}^{\ast})$ as $\delta^{\ast}(X)$. Likewise for the other models.
	
	It suffices to show that ${\Delta}_{mr}^*$ has expectation $\Delta$ in the union model, where 
	$$
	{\Delta}_{mr}^* = \mathbb{P}_n \left\{  \dfrac{1}{\delta^D(X;{\beta}^*_{\text{\it dr}})}\left(Y-D\delta \left( X; {\alpha}^*_{\text{\it dr}} \right) -p_{0}^Y \left( X; {\iota}^*_{\text{\it mle}} \right) +p_{0}^{D}(X; {\theta}^*_{\text{\it mle}}) \delta \left(X; {\alpha}^*_{\text{\it dr}} \right)   \right)  \dfrac{2Z-1 }{f(Z\mid X; {\gamma}^*_{\text{\it mle}}) } +\delta \left( X;{\alpha}^*_{\text{\it dr}} \right) \right\}.
	$$
	
	Suppose only  $\m_1$ holds such that $\delta^{\ast} \left( X\right) = \delta \left( X\right)$, $\delta^{D,\ast}(X)=\delta^D(X), p_{0}^{Y,\ast}
	\left( X\right) = p_{0}^Y
	\left( X\right)$ and $p_{0}^{D,\ast}(X) =p_{0}^{D}(X)$ but $f^{\ast}\left( Z\mid X\right)\neq f(Z\mid X)$. Then
	\begin{eqnarray*}
		E[{\Delta}_{\text \it mr}^*]	&=&E\left\{ \dfrac{2Z-1}{f^{\ast }\left( Z\mid X\right) }
		\dfrac{1}{\delta^D(X)}    
		\left\{ {Y}-{p_{0}^Y \left( X\right) }-{D}\delta \left(
		X\right)  +p_{0}^{D}(X) {\delta \left(
			X\right) } \right\}+\delta \left( X\right)  
		\right\}  \\
		&=&E\left\{ \dfrac{2Z-1}{f^{\ast }\left( Z\mid X\right) }%
		\dfrac{1}{\delta^D(X)}    
		\left\{
		{\delta \left( X\right) \delta^D(X)Z }
		-{Z\delta^D(X) }\delta \left(
		X\right) \right\} + \Delta
		\right\}  \\
		&=&\Delta.
	\end{eqnarray*}%
	Suppose instead only $\m_2$ holds such that 
	$f^\ast \left( Z\mid X\right) = f(Z\mid X) $ and $\delta^{D,\ast} \left(
	X\right) = \delta^D(X)$ but $\delta^{\ast} \left( X\right) \neq \delta \left( X\right), p_{0}^{Y,\ast}
	\left( X\right) \neq p_{0}^Y
	\left( X\right)$ and $p_{0}^{D,\ast}(X) \neq p_{0}^{D}(X)$.
	Then 
	\begin{small}
		\begin{flalign*}
		E[{\Delta}^*_{\text \it mr}]	&\hspace*{-0.0cm}= 	E\left\{ \dfrac{2Z-1 }{f\left( Z\mid X\right) }
		\dfrac{1}{\delta^D(X)}
		\left( 
		{Y}-{%
			p_{0}^{Y,\ast }\left( X\right) }-{D}\delta ^{\ast }\left( X\right)  
		+p_{0}^{D,\ast}\left( X\right) {\delta
			^{\ast }\left( X\right) }		\right)+\delta ^{\ast }\left(
		X\right)   \right\}  \\
		&\hspace*{-0.0cm}= E\left\{ \dfrac{2Z-1}{f\left( Z\mid X\right) } \dfrac{1}{\delta^D(X)}\left( 
		{\delta \left( X\right) \delta^D(X)Z + p_0^Y(X) }-{p_{0}^{Y,\ast }\left( X\right) }-\left(\delta^D(X) Z + p_{0}^{D}(X) \right)\delta
		^{\ast }\left( X\right)  
		+p_{0}^{D,\ast}\left( X\right) {%
			\delta ^{\ast }\left( X\right) }\right) \right\} +  \\   &\hspace*{-0.0cm} \quad E\left\{ \delta ^{\ast}\left( X\right)   \right\}  \\[3pt]
		&\hspace*{-0.0cm}= E\left\{ \delta \left( X\right) -\delta ^{\ast }\left( X\right)
		+\delta ^{\ast }\left( X\right)   \right\} =\Delta.
		\end{flalign*}%
	\end{small}
	Finally suppose only $\m_3$ holds such that $f^\ast\left( Z\mid X\right) =f(Z\mid X)$ and $\delta^*
	\left( X\right) = \delta(X) $ but  $\delta^{D,\ast}(X)\neq \delta^D(X), p_{0}^{Y,\ast}
	\left( X\right) \neq  p_{0}^Y
	\left( X\right)$ and $p_{0}^{D,\ast}(X) \neq p_{0}^{D}(X)$. Then 
	\begin{small}
		\begin{flalign*}
		E[{\Delta}^*_{\text \it mr}]	&= E\left\{ \dfrac{2Z-1}{f\left( Z\mid X\right) } \dfrac{1}{\delta^{D,\ast}(X) } 
		\left( 
		{Y}-{%
			p_{0}^{Y,\ast }\left( X\right) }
		-{D}\delta \left( X\right)  +p_{0}^{D,\ast}\left( X\right) {\delta \left( X\right) }\right)+\delta \left( X\right)  
		\right\}  \\
		&= E\left\{ \dfrac{2Z-1}{f\left( Z\mid X\right) } \dfrac{1}{\delta^{D,\ast}(X) }  \left( 
		{\delta \left( X\right) \delta^D(X) Z + p_0^Y(X)}
		-{p_{0}^{Y,\ast }\left( X\right) }-\left(\delta^D(X) Z + p_0^D(X)\right)\delta \left( X\right)  	 +p_{0}^{D,\ast}\left( X\right) {\delta
			\left( X\right) }%
		\right) \right\} + \\
		&\quad  E\left\{\delta \left( X\right)   \right\}  \\
		&= E\left\{ \dfrac{1}{\delta^{D,\ast}\left(
			X\right) }
		\left( {\delta \left( X\right) \delta^D(X) } -{\delta \left( X\right) \delta^D
			\left( X\right) } \right)+\delta \left( X\right) 
		\right\} =\Delta.
		\end{flalign*}
	\end{small}
	This concludes the proof for the first claim. The second claim in Theorem 6 follows from Theorem 5.
	
	\section{Additional simulation results}

	\begin{table}
		% 1.6 Summary.R - submitted050317
		\begin{center}
			\small
			\caption{Boxplots for the proposed estimators under 
				all possible combinations of correct/incorrect specifications of the following four sets of models: $\delta(X), \delta_D(X), f(Z\mid X)$ and $OP(X)$, where $OP(X) = OP^Y(X), OP^D(X))$.  The true value for the average Wald estimand is 0.087, annotated with the red horizontal lines. The sample size is 500}
			\bigskip
			\label{tab:est2}
			\begin{tabular}{c m{2cm}ccccccccc}
				\toprule
				&\multirow{2}{*}	{$(\pi(X),OP(X))$ } &&  \multicolumn{4}{c}	{$(\delta^D(X),\delta(X))$}   \\[2pt]
				\cmidrule(l){4-7} 
				&	 && $(\checkmark,\checkmark)$ & $(\checkmark,\times)$   & $(\times,\checkmark)$ & $(\times,\times)$   \\
				\midrule
				&$(\checkmark,\checkmark)$   && \adjustimage{height=.22\textwidth,valign=m}{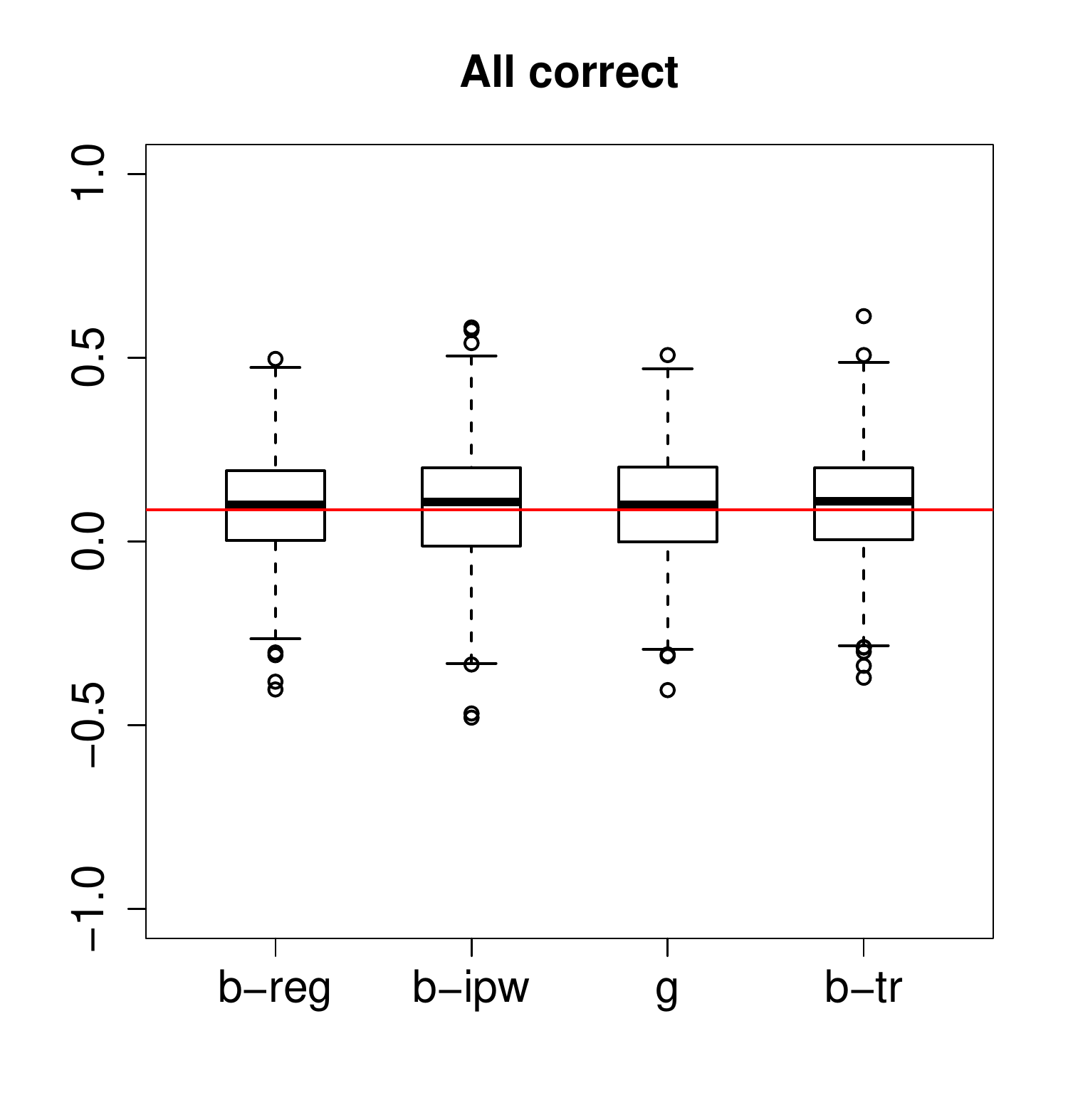} & \adjustimage{height=.22\textwidth,valign=m}{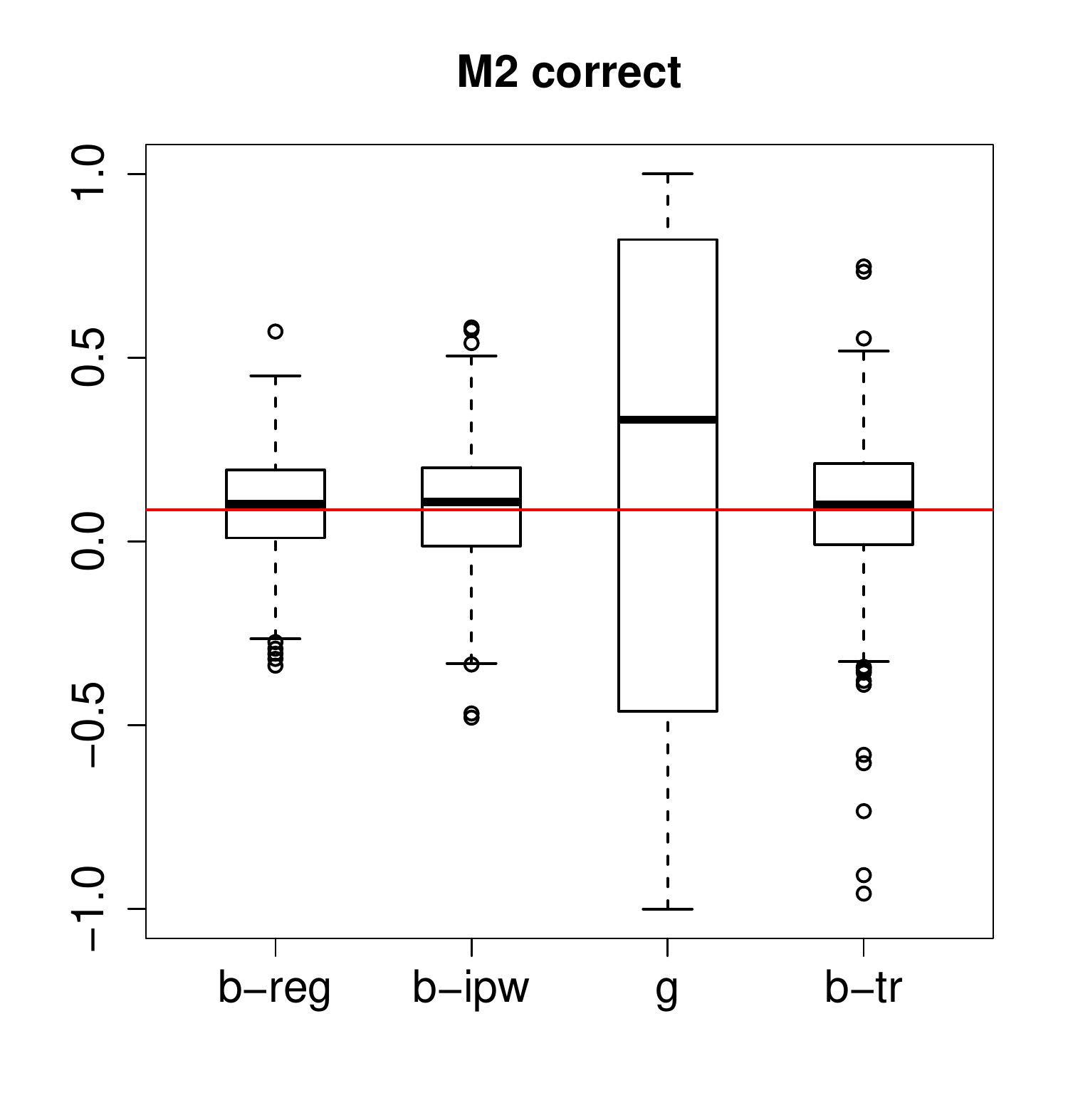} & \adjustimage{height=.22\textwidth,valign=m}{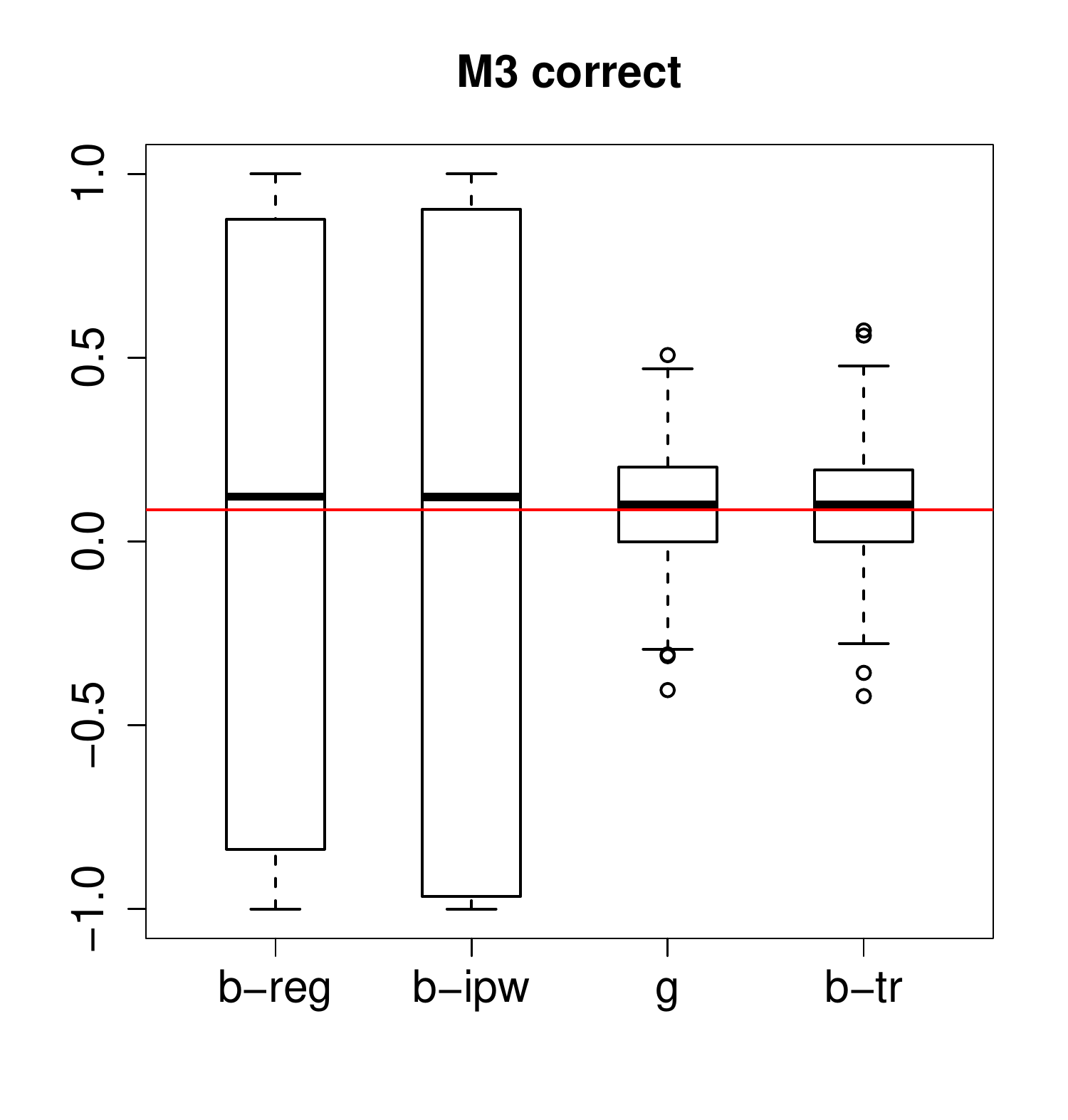}& \adjustimage{height=.22\textwidth,valign=m}{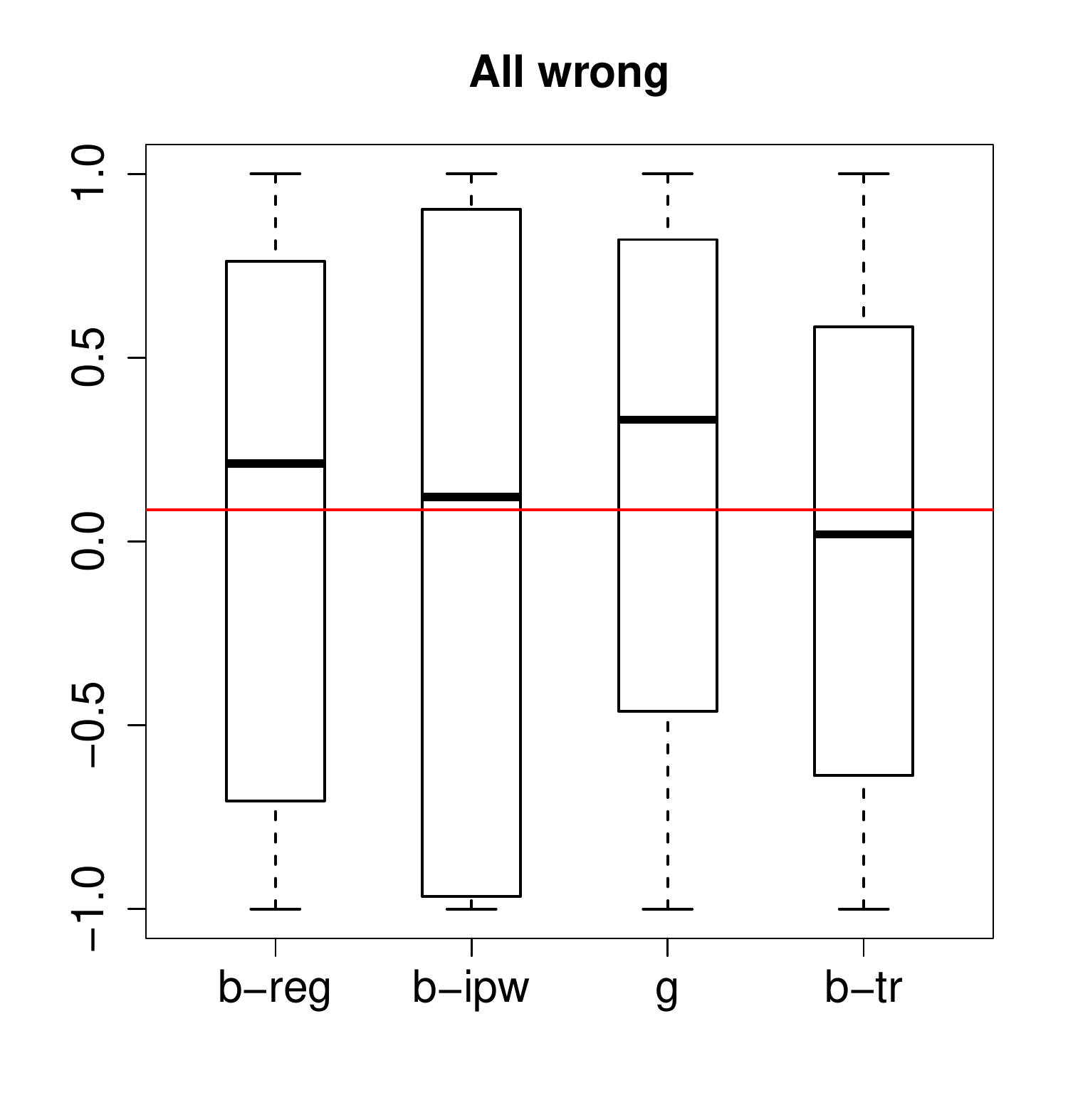}  \\ [5pt]
				&$(\checkmark,\times)$  && \adjustimage{height=.22\textwidth,valign=m}{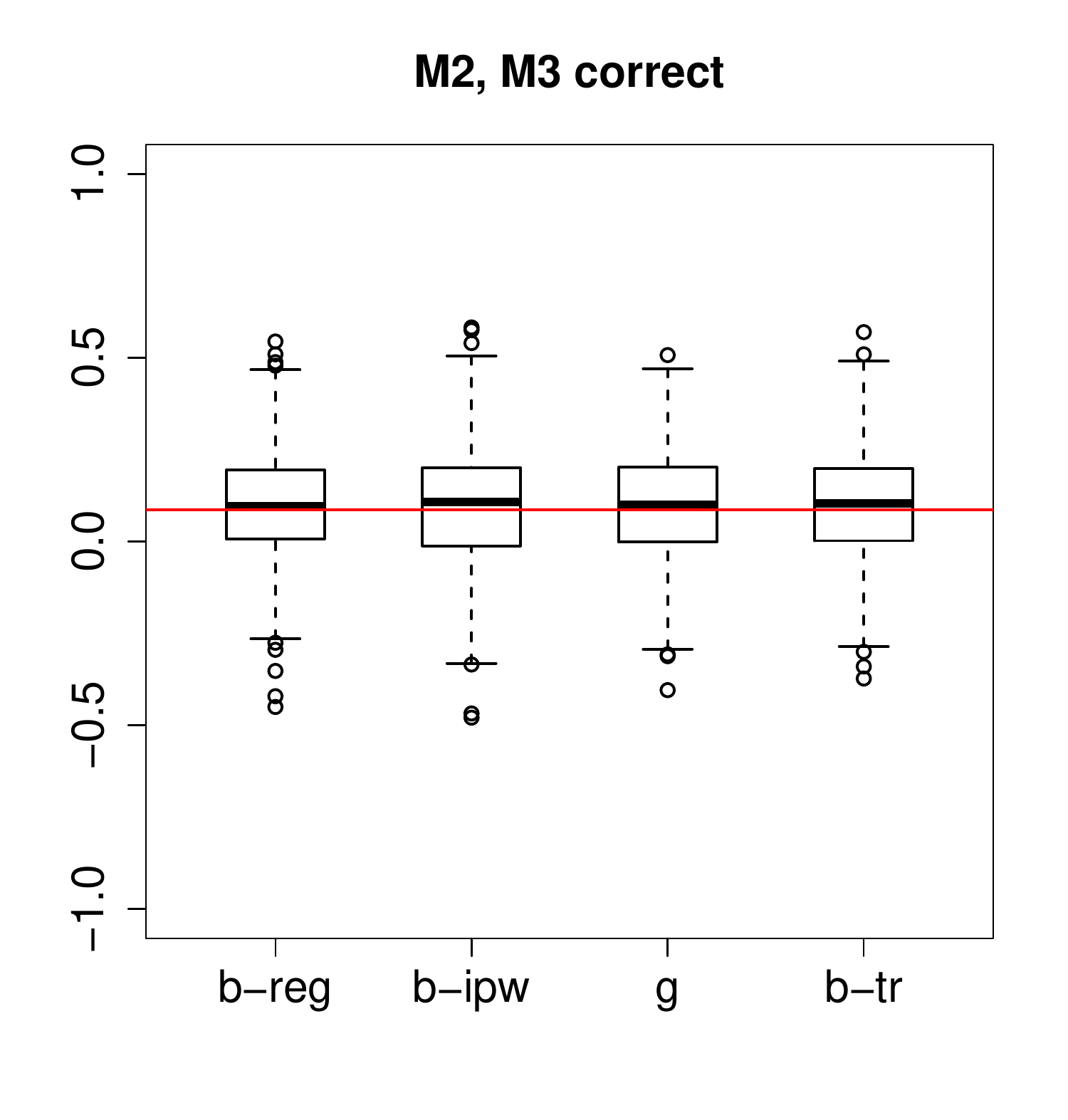} & \adjustimage{height=.22\textwidth,valign=m}{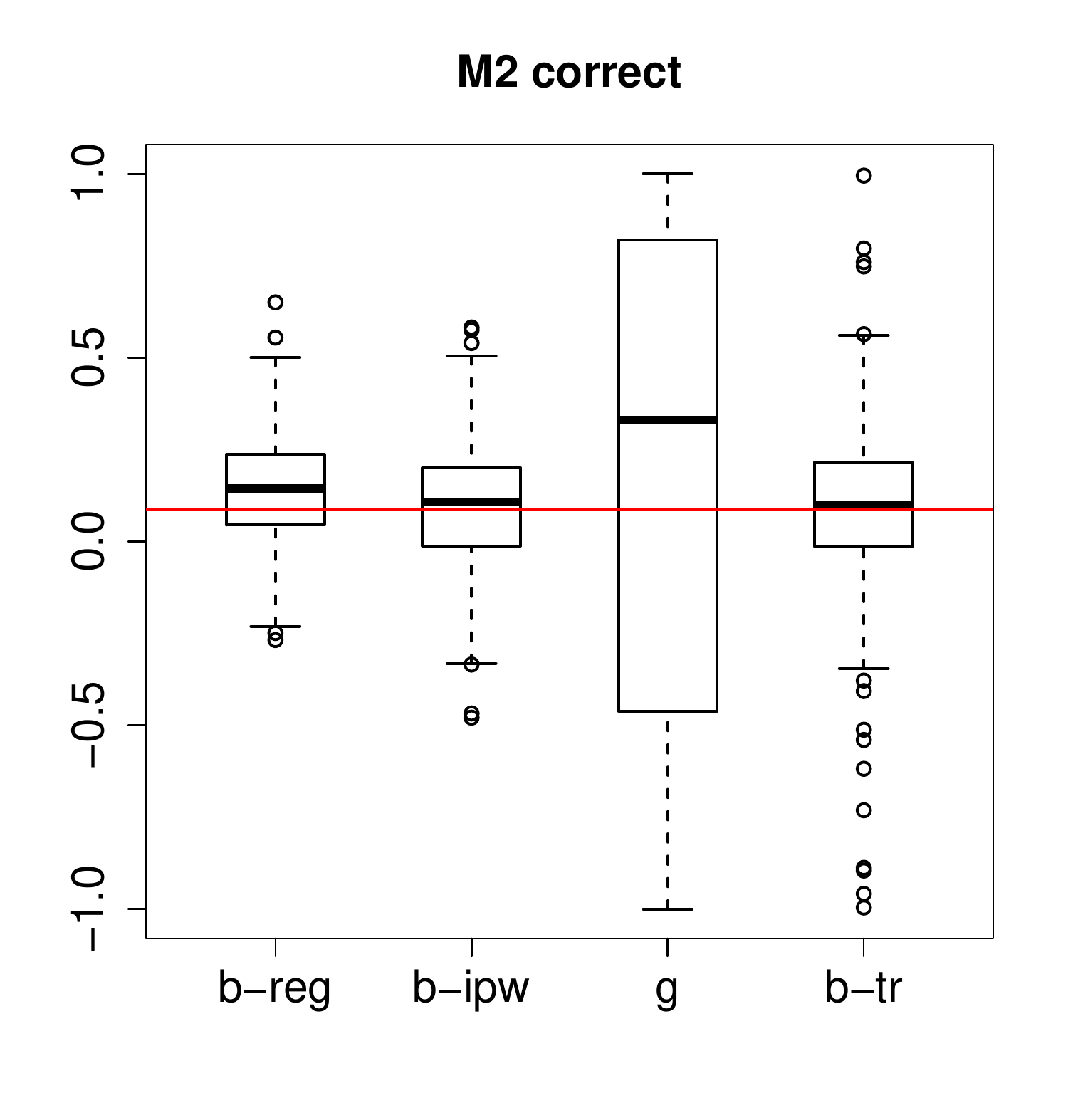} & \adjustimage{height=.22\textwidth,valign=m}{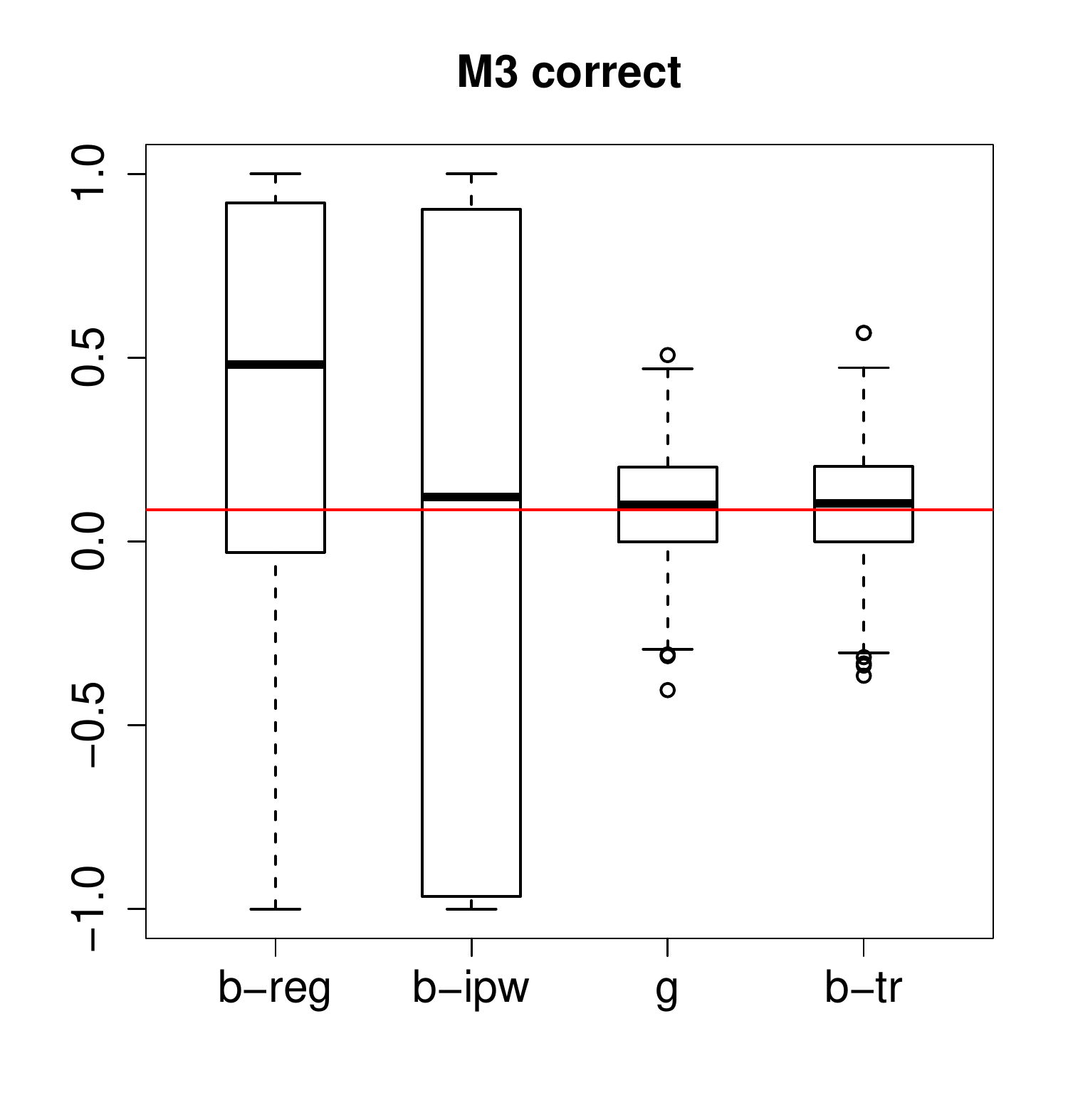}& \adjustimage{height=.22\textwidth,valign=m}{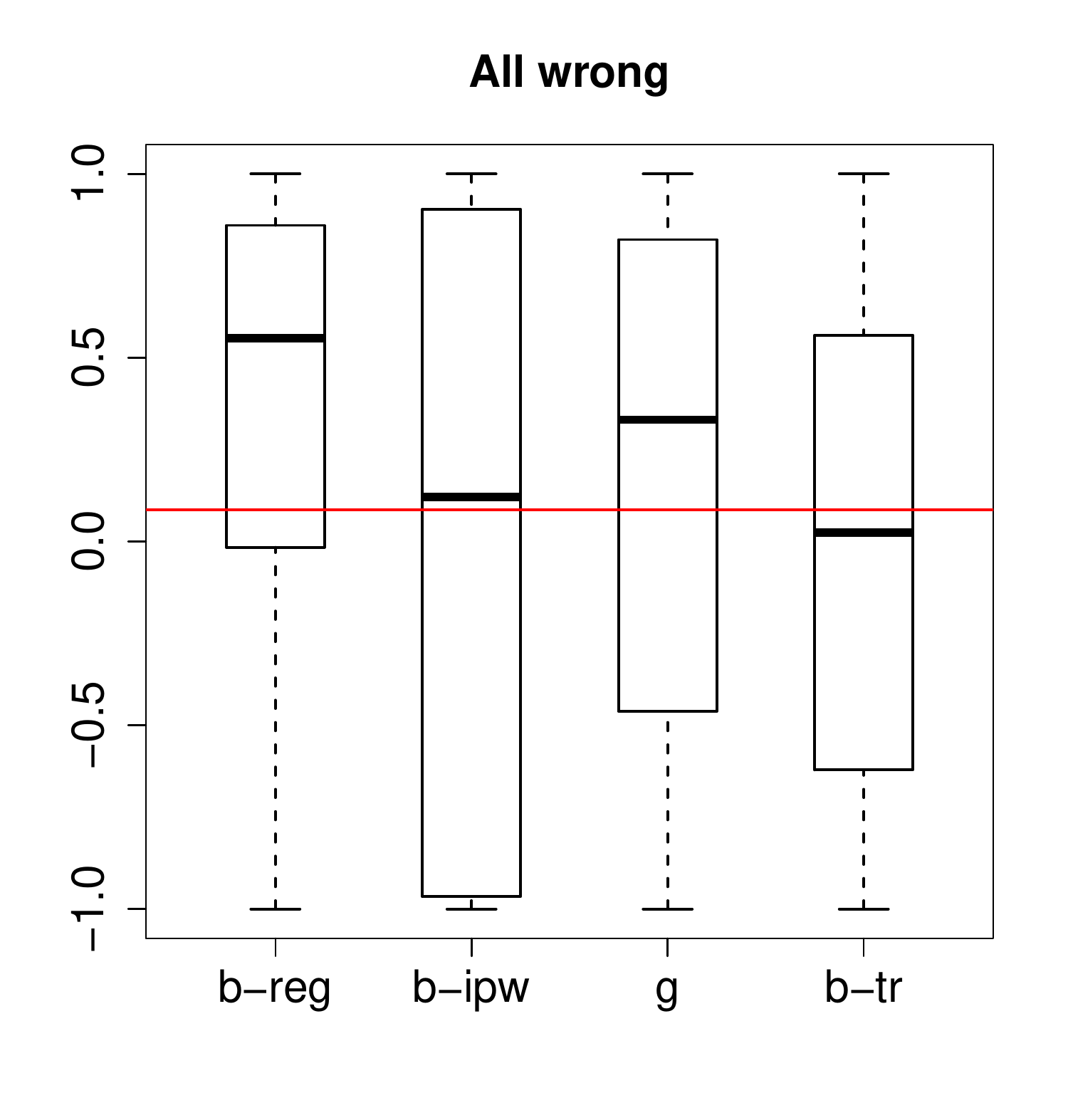}  \\ [5pt]
				&$(\times,\checkmark)$   && \adjustimage{height=.22\textwidth,valign=m}{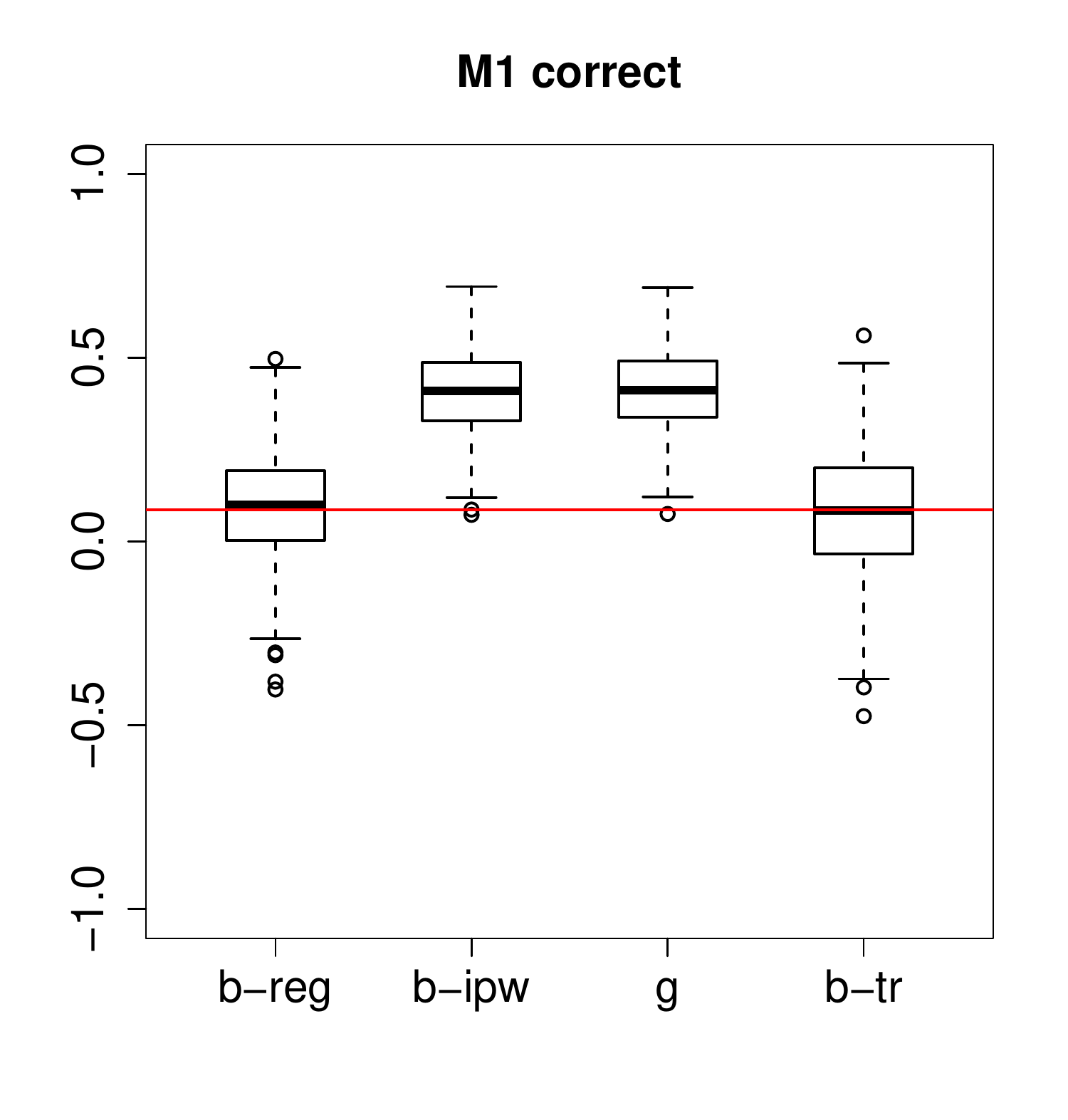} & \adjustimage{height=.22\textwidth,valign=m}{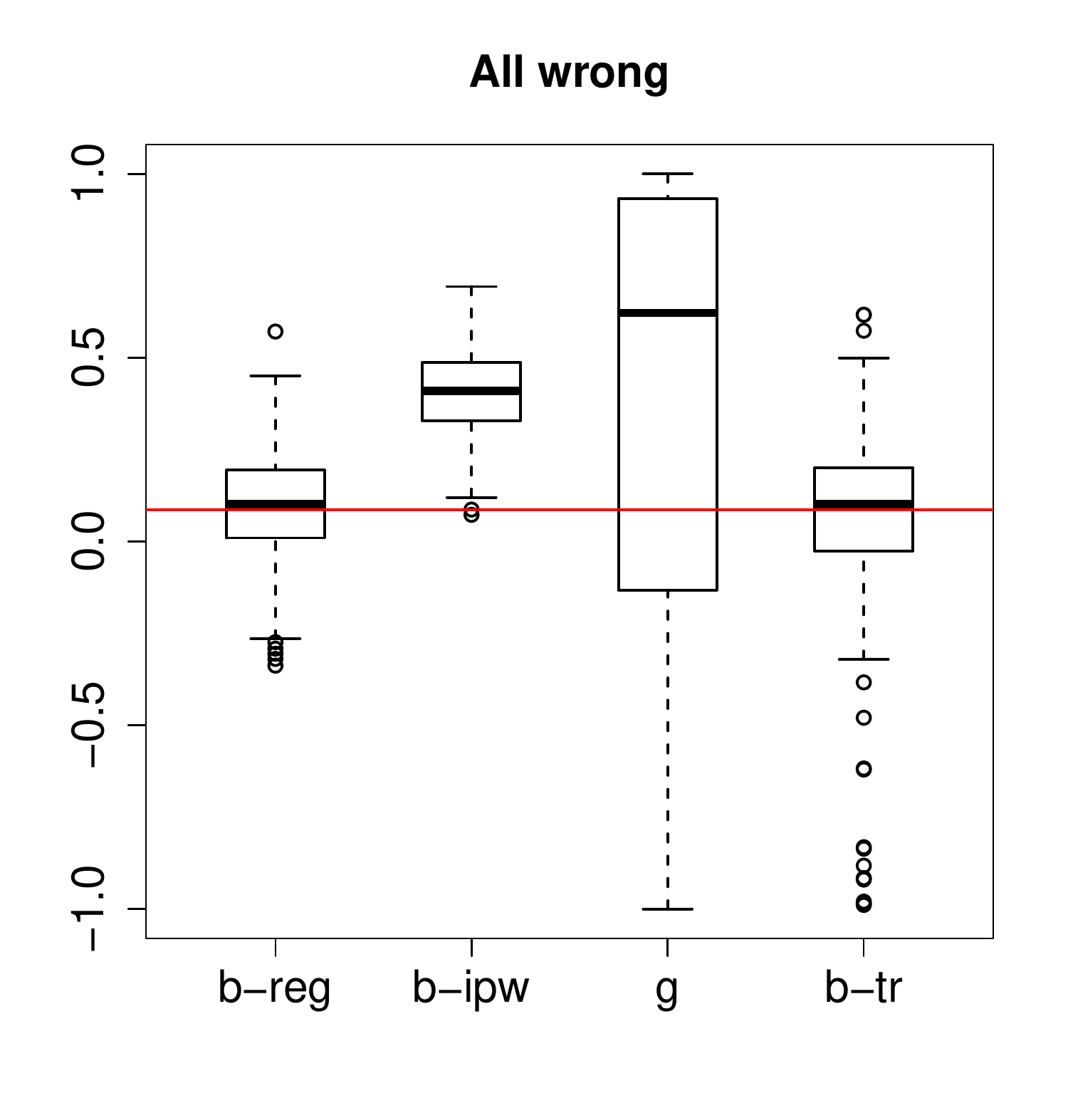} & \adjustimage{height=.22\textwidth,valign=m}{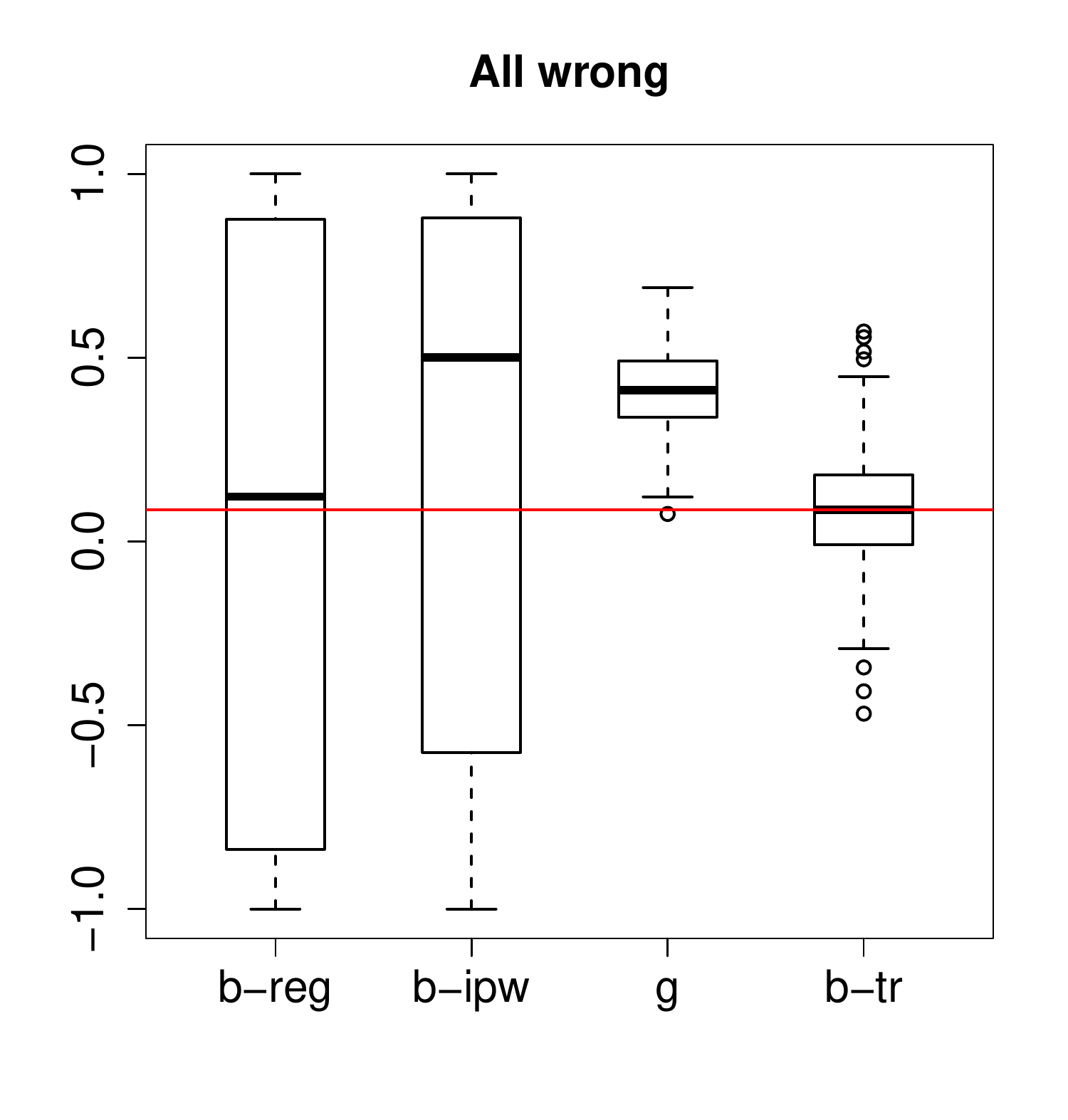}& \adjustimage{height=.22\textwidth,valign=m}{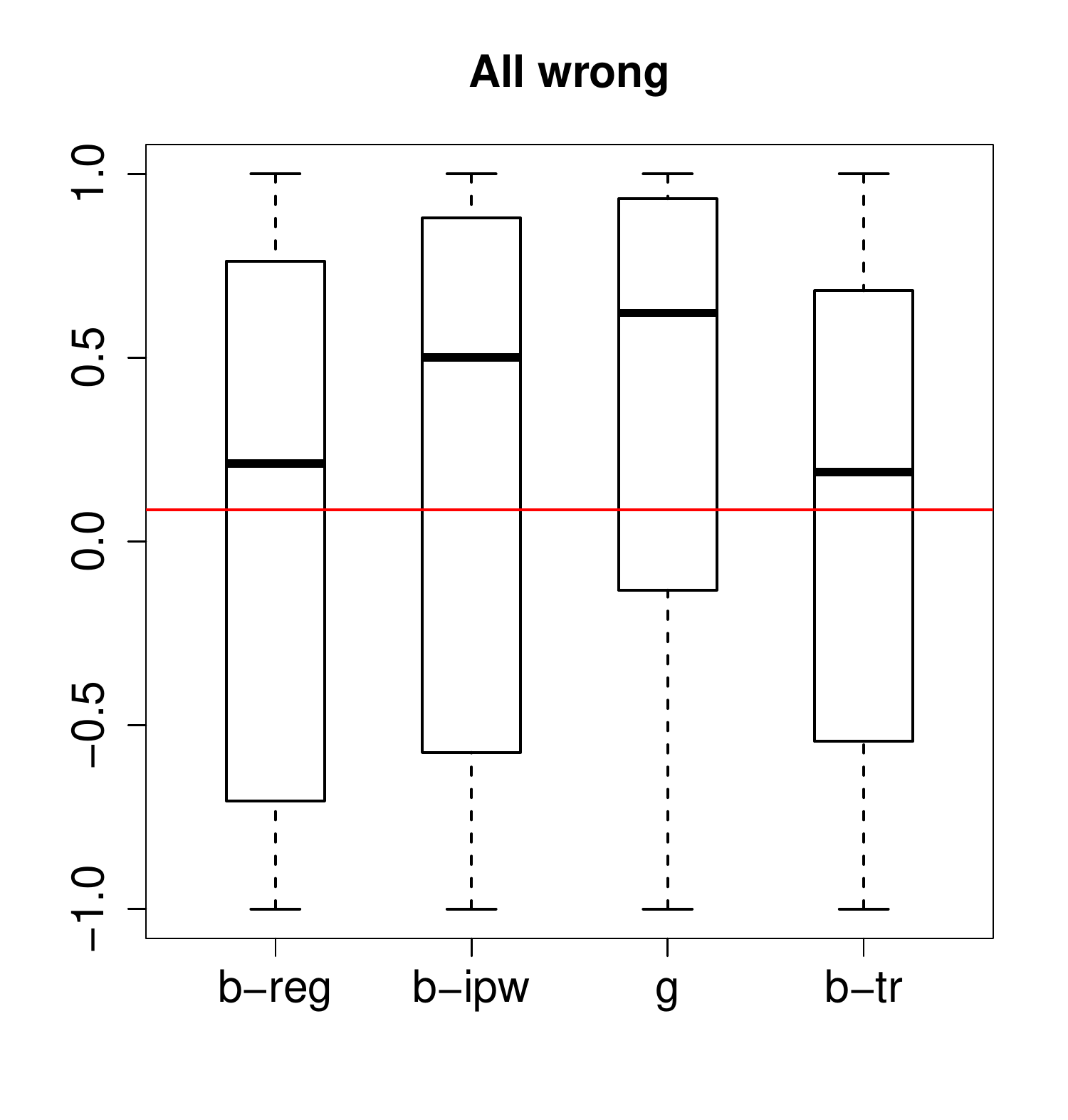}  \\ [5pt]
				&$(\times,\times)$  && \adjustimage{height=.22\textwidth,valign=m}{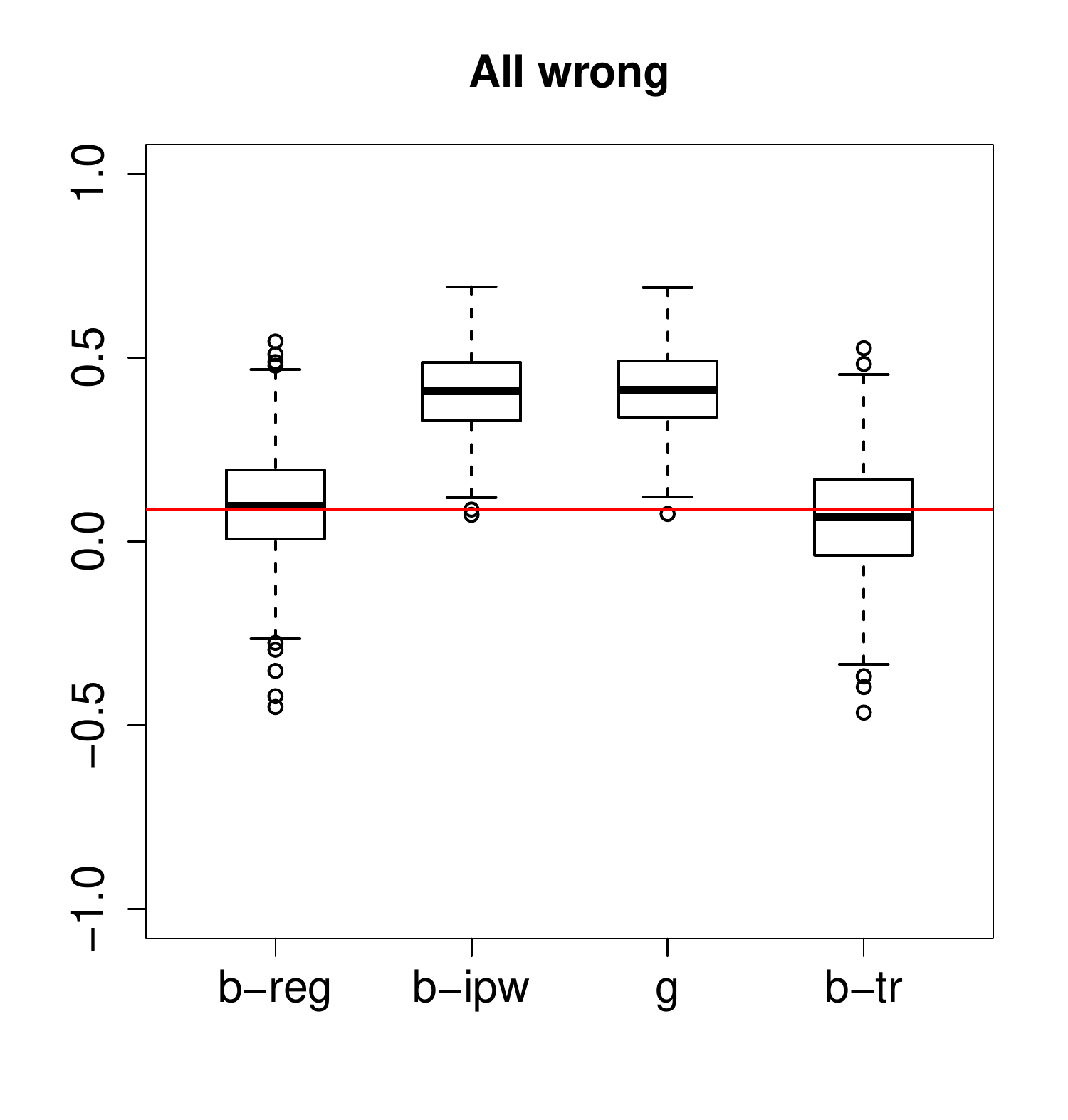} & \adjustimage{height=.22\textwidth,valign=m}{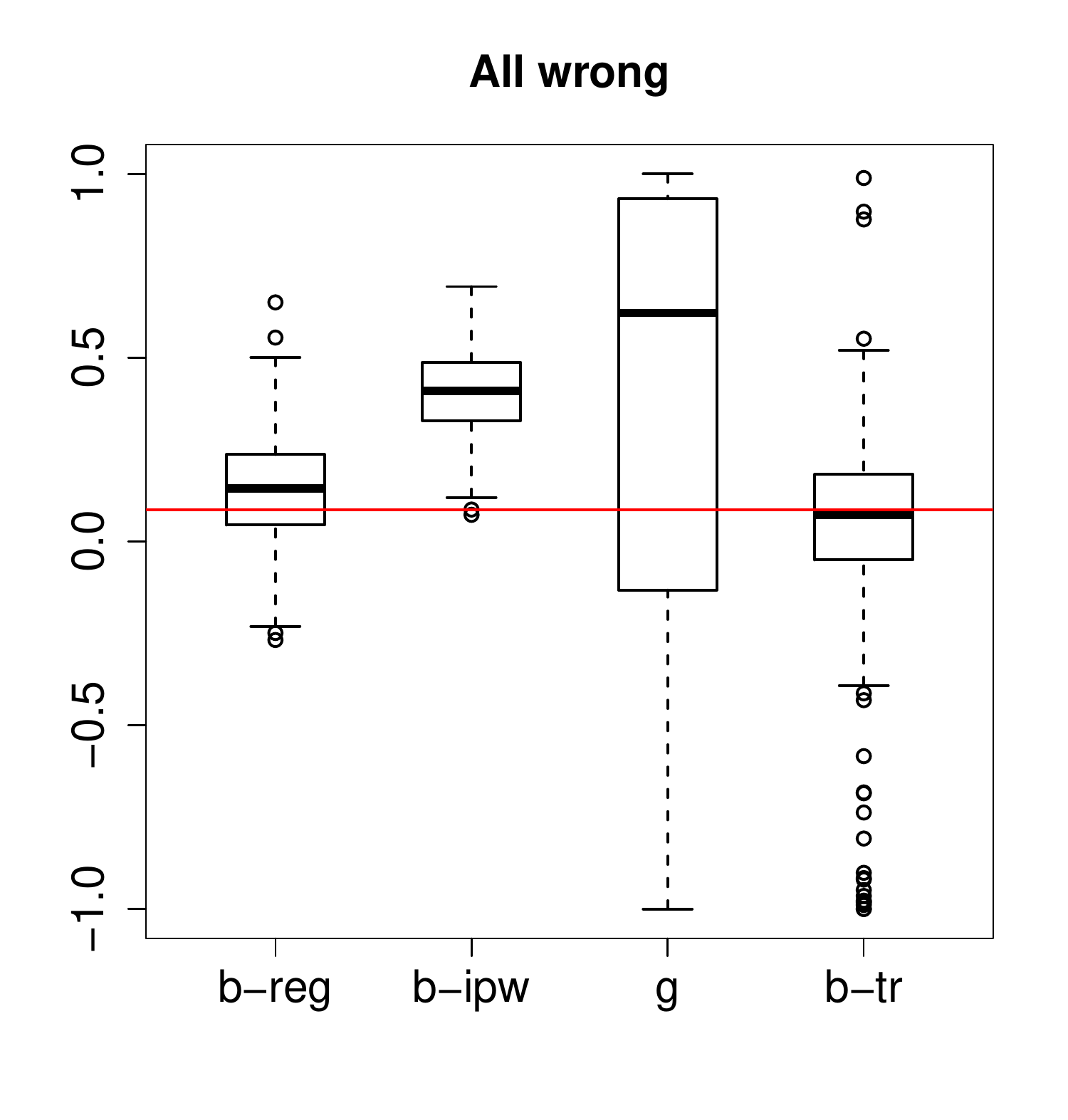} & \adjustimage{height=.22\textwidth,valign=m}{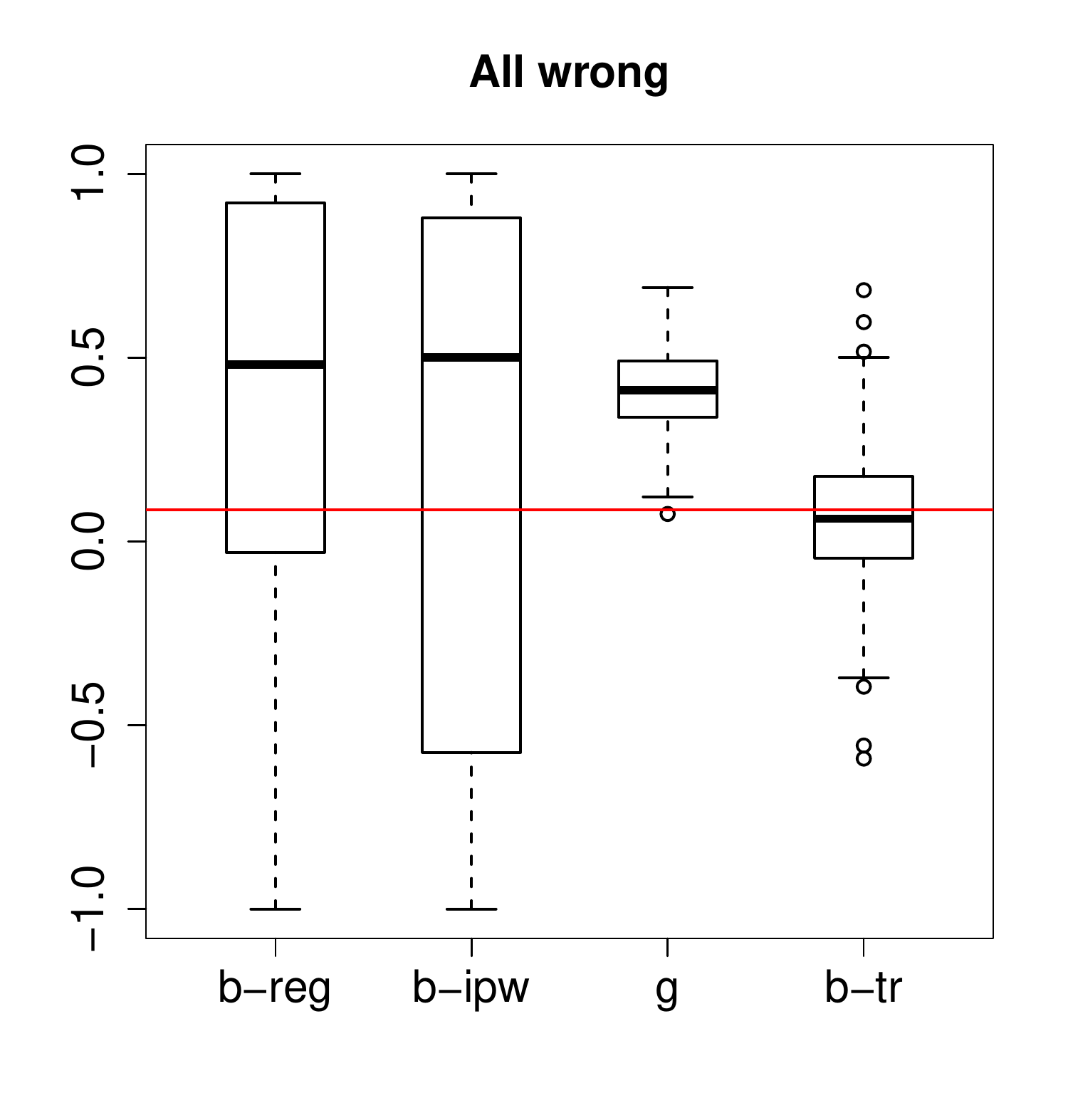}& \adjustimage{height=.22\textwidth,valign=m}{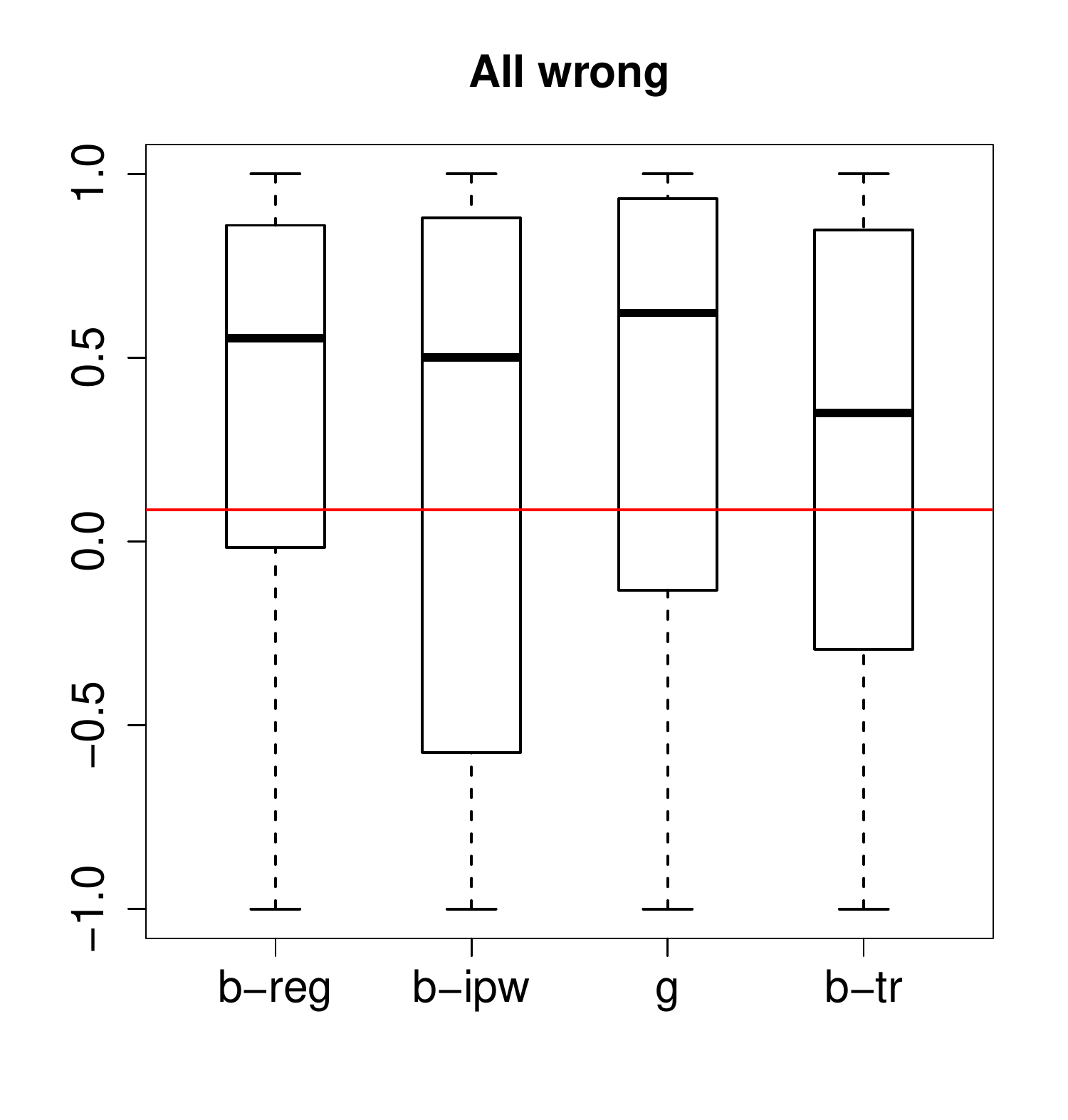}  \\ 
				\bottomrule 
			\end{tabular}
		\end{center}
	\end{table}

\end{document}